%% file: Main.tex
\documentclass[orbibl,a4paper]{LMCS}

\def\doi{7 (4:09) 2011}
\lmcsheading%
{\doi}
{1--23}
{}
{}
{Nov.~29, 2009}
{Dec.~21, 2011}
{}

\usepackage{hyperref} 
\usepackage{paralist,amssymb,xspace,manfnt,url,bcprules}
\usepackage[usenames]{color}
\usepackage{qtree}

\newif\iffull\fulltrue
\newif\ifdraft\draftfalse

\newcommand\eop{$\Box$}
\newtheorem{proposition}[thm]{Proposition}
\newtheorem{remark}{Remark}
\newtheorem{ex}{Example}
\newenvironment{example}{\begin{ex}\rm}{\hfill \eop\end{ex}}

\ifdraft
\newcommand\todo[1]{{\footnotesize \color{OliveGreen}
[\textbf{To do:} #1]}}
\newcommand\nk[1]{{\footnotesize \color{blue}
[#1 - \textbf{Naoki}]}}
\newcommand\lo[1]{{\footnotesize \color{red}
[#1 - \textbf{Luke}]}}
\newcommand\changed[1]{{\color{blue}{#1}}}
\newcommand\lochanged[1]{{\color{red}{#1}}}
\newcommand\review[1]{{\footnotesize \color{green}
[Reviewer says: #1]}}
\else
\newcommand\todo[1]{}
\newcommand\nk[1]{}
\newcommand\lo[1]{}
\newcommand\review[1]{}
\newcommand\changed[1]{#1}
\newcommand\lochanged[1]{#1}
\fi

\title[Complexity of Model Checking Recursion Schemes]{Complexity of Model Checking Recursion Schemes for Fragments of the Modal Mu-Calculus}

\author[N.~Kobayashi]{Naoki Kobayashi}	
\address{Graduate School of Information Sciences, Tohoku University, 6-3-09 Aoba, Aramaki, Aoba-ku Sendai, 980-8579 Japan}	 
\urladdr{\url{koba@ecei.tohoku.ac.jp}}  

\author[C.-H.~L.~Ong]{C.-H.~Luke Ong}
\address{Department of Computer Science, University of Oxford, Wolfson Building, Parks Road, Oxford OX1 3QD, UK}
\urladdr{\url{Luke.Ong@cs.ox.ac.uk}}

\input{local}

\newcommand\cal{\mathcal}

\begin{document}
\input{abstract}

\keywords{model checking, higher-order recursion schemes, modal
mu-calculus, complexity}
\subjclass{F.3.1, D.2.4}
\maketitle
\input{intro}
\input{pre}
\input{trivialAPT}

\input{disjunctiveAPT}
\input{app}

\iffull
\input{related}

\fi
\input{conc}
\input{ack}


\bibliographystyle{abbrv}
\bibliography{full,koba,ong89-90}

\iffull
\newpage
\appendix
\input{Proof-EquiExpressivity}
\input{upperbound-disjunctiveAPT}
\fi
\end{document}

%% file: local.tex
\newcommand{\EPSILON}{\texttt{Epsilon}}
\newcommand{\READ}{\texttt{Read}}
\newcommand{\ACCEPT}{\texttt{Accept}}
\newcommand\kinds{types}
\newcommand\kind{type}
\newcommand\AtomTy[1]{\mathcal{T}(#1)}
\newcommand\EAT{\theta}
\newcommand\stack{s}
\newcommand\redlab[1]{\stackrel{#1}{\mathbin{\,\rightarrow\,}}}
\newcommand\topn[1]{{\mathit top}_{#1}}

\newcommand\stacktop{\topn{1}}

\newcommand\Ta{\terminal{a}}
\newcommand\Tb{\terminal{b}}
\newcommand\Tc{\terminal{c}}
\newcommand\Tr{\terminal{r}}
\newcommand\Te{\terminal{e}}
\newcommand\WA{\mathcal{C}} 
\newcommand\APT{\mathcal{A}}
\newcommand\dAPT{\mathcal{D}}
\newcommand\Pfun{\Omega}
\newcommand\M{\mathcal{M}}
\newcommand\lang{\mathcal{L}}

\newcommand\APDA{\mathcal{A}}

\newcommand\trPDA[2]{\mathcal{M}_{#1,#2}}
\newcommand\tPDA{\mathcal{M}}
\newcommand\trBUCHI[2]{\mathcal{B}_{#2}}
\newcommand\UNIV{\mathtt{A}}
\newcommand\EXISTS{\mathtt{E}}
\newcommand\TOP{\mathtt{T}}
\newcommand\BOT{\mathtt{R}}
\newcommand\OP[1]{\mathit{Op}_{#1}}
\newcommand\ID{\mathtt{id}}

\newcommand\terminal[1]{\mathtt{#1}}
\newcommand\order{\mathit{order}}
\newcommand\arity{\mathit{arity}}
\newcommand\TERMS{\Sigma}
\newcommand\NONTERMS{\mathcal{N}}
\newcommand\RULES{\mathcal{R}}
\newcommand\Hra{\ra}
\newcommand\Hred[1]{\red{#1}}
\newcommand\Hreds[1]{\reds{#1}}
\newcommand\GRAM{\mathcal{G}}
\newcommand\Bot[1]{{#1}^{\bot}}

\newcommand\IT{\bigwedge}

\newcommand\unit{\star}

\newcommand\spec{\changed{q}}
\newcommand\NU[1]{\mathbf{\nu}^{#1}}
\newcommand\NEW[1]{\mathbf{New}^{#1}}
\newcommand\nuexp[1]{\mathbf{New}^{#1}\ }
\newcommand\ACC[1]{\mathbf{Acc}_{#1}}
\newcommand\acc[3]{\ACC{#1}\ #2\ #3}

\newcommand\IFNONDET{\mathbf{If*}}
\newcommand\ifexp[1]{\IFNONDET\ {#1}\ } 
\newcommand\Tres{\mathbf{R}}
\newcommand\Tvoid{\mathbf{unit}}

\newcommand\f{{a}}  

\newcommand\ERROR{\mathbf{Error}}
\newcommand\rst{\rho}

\newcommand\red[1]{\longrightarrow_{#1}}
\newcommand\reds[1]{\red{#1}^*}

\newcommand\BR{\mathtt{br}}

\newcommand\T{\mathtt{o}}  
\newcommand\sem[1]{\mathbin{[\![}#1\mathbin{]\!]}}

\newcommand\TRUE{{\sf t}}
\newcommand\FALSE{{\sf f}}

\newcommand\subst[2]{[#2/#1]}
\newcommand\ra{\rightarrow}
\newcommand\set[1]{\{#1\}}

\newcommand\IFF{\Longleftrightarrow}

\newcommand\seq[1]{\widetilde{#1}}

\newcommand\dom{\mathit{dom}}
\newcommand\COL{\mathbin{:}}

\newcommand\MAX{\mathit{max}}

\newcommand\gtop[1]{\textit{g2p}(#1)}

%% file: abstract.tex
\begin{abstract}

Ong has shown that the modal mu-calculus model checking problem
(equivalently, the alternating parity tree automaton (APT) acceptance
problem) of possibly-infinite ranked trees generated by order-$n$
recursion schemes is \(n\)-EXPTIME complete. We consider two
subclasses of APT and investigate the complexity of the respective
acceptance problems. The main results are that, for APT with a single
priority, the problem is still \(n\)-EXPTIME complete;
whereas, for APT with a disjunctive transition
function, the problem is \((n-1)\)-EXPTIME complete. This study was
motivated by Kobayashi's recent work showing that the resource usage
verification of functional programs can be reduced to the model
checking of recursion schemes. As an application, we show that the
resource usage verification problem is \((n-1)\)-EXPTIME complete.

\end{abstract}

%% file: intro.tex
\section{Introduction}
\label{sec:intro}

The model checking problem for higher-order recursion schemes has been
a topic of active research in recent years (for motivation as to why
the problem is interesting, see e.g.~the introduction of Ong's paper
\cite{Ong06LICS}).
This paper studies the complexity of the problem with respect to certain fragments
of the modal \(\mu\)-calculus.
A higher-order recursion scheme  (recursion scheme, for short) is a kind of (deterministic) grammar for generating
a possibly-infinite ranked tree.  The model checking problem for
recursion schemes is to decide, given an order-$n$ recursion scheme
\(\GRAM\) and a specification \(\psi\) for infinite trees, whether the
tree generated by \(\GRAM\) satisfies \(\psi\).  Ong~\cite{Ong06LICS}
has shown that if \(\psi\) is a modal \(\mu\)-calculus formula (or
equivalently, an alternating parity tree automaton), then the model
checking problem is \(n\)-EXPTIME complete.

Following Ong's work, Kobayashi~\cite{Kobayashi09POPL} has recently applied the decidability result
to the model checking of higher-order functional programs
(precisely, programs of the
simply-typed \(\lambda\)-calculus with recursion
and resource creation/access primitives).
He considered
the \emph{resource usage verification problem}~\cite{IK05TOPLAS}---the
problem of whether programs access dynamically created
resources in a valid manner (e.g.~whether every opened file will
eventually be closed, and thereafter never read from or written to before it is reopened).
He showed that the resource usage verification problem reduces to
a model checking problem for recursion schemes by
giving a transformation that, given a functional program,
constructs a recursion scheme that generates all possible resource access sequences of the program.
From Ong's result, it follows that the resource usage verification problem
is in \(n\)-EXPTIME (where, roughly, \(n\) is the highest order of types in the program). This result also implies
that various other verification problems,
including ({the} precise verification of)
reachability (``{Does a closed program} reach the fail command?'')
and flow analysis (``Does a sub-term \(e\) evaluate
to a value generated at program point \(l\)?''), are also in \(n\)-EXPTIME,
as they can be easily recast as resource usage verification problems.

It was however unknown whether \(n\)-EXPTIME is the {tightest} upper-bound of the resource usage verification problem.
Although the model checking of recursion schemes is \(n\)-EXPTIME-hard
for the full modal \(\mu\)-calculus, only a certain fragment of the modal \(\mu\)-calculus is used in Kobayashi's approach to the resource usage verification problem.
First, specifications are restricted to safety properties, which can be described by
B\"u{chi} tree automata with a trivial {acceptance} condition
(the class called ``trivial automata'' by Aehlig~\cite{Aehlig07}). Secondly, specifications are
also restricted to linear-time properties---the branching structure of
trees is ignored, and only the path languages of trees are of interest.
Thus, one may reasonably hope that there is a more tractable model checking algorithm
than the \(n\)-EXPTIME algorithm.

The goal of this paper is, therefore, to study the complexity of the model checking of
recursion schemes for various fragments of the modal \(\mu\)-calculus (or, alternating
parity tree automata) and to apply the result to obtain tighter
bounds of the complexity  of the resource usage verification problem.

The main results of this paper are as follows:
\begin{asparaenum}[(i)]
\item The problem of whether a given B\"{u}chi {tree} automaton with a trivial acceptance condition (or, equivalently, alternating parity tree automaton with a single priority \(0\))
accepts the tree generated by an order-$n$ recursion scheme is still \(n\)-EXPTIME-hard,
both in the size of the recursion scheme and that of the automaton.
This follows from the \(n\)-EXPTIME-completeness
of {the word acceptance problem} of higher-order {alternating}
pushdown automata\footnote{Engelfriet's proof \cite{Engelfriet91} is
for a somewhat different (but equivalent) machine which is called \emph{iterated pushdown automaton}.}
~\cite{Engelfriet91}. 

\item We introduce a new subclass of alternating parity tree automata (APT) called
\emph{disjunctive APT}, and show that its acceptance problem for trees generated by order-$n$ recursion schemes is \((n-1)\)-EXPTIME complete.
From this general result, it follows that both the linear-time properties
(including reachability, which is actually
\((n-1)\)-EXPTIME-complete) and finiteness of the tree generated by a recursion scheme are \((n-1)\)-EXPTIME.
\item As an application, we show that
the resource usage verification problem~\cite{Kobayashi09POPL} is also
(\(n-1\))-EXPTIME-complete, where \(n\) is the highest order of types
used in the source program (written {in an appropriate language \cite{Kobayashi09POPL}}).
\end{asparaenum}
\iffull

The rest of this section is organized as follows.
Section~\ref{sec:pre} reviews definitions of recursion schemes and alternating parity tree automata (APT).
Section~\ref{sec:ta-hardness} introduces the class of trivial APT and studies the complexity of model checking
recursion schemes.
Section~\ref{sec:disjunctiveAPT} introduces the class of disjunctive APT and studies the complexity of model checking
recursion schemes.
Section~\ref{sec:app} applies the result to analyze the complexity of the resource usage verification.
Section~\ref{sec:related} discusses related work and concludes the paper.
\fi

%% file: pre.tex
\section{Preliminaries}
\label{sec:pre}

\newcommand\N{\mathbb{N}}


Let \(\Sigma\) be a ranked alphabet, i.e. a function that
  maps a terminal symbol to its arity, which is a non-negative
  integer. Let \(\N = \set{1, 2, \cdots }\). A \(\Sigma\)-labeled
(unranked) tree \(T\) is a partial map from \(\N^*\) to
\(\dom(\Sigma)\), such that \(s \, k\in \dom(T)\) (where \(s\in \N^*,
k\in \N\)) implies \(\set{s}\cup \set{s j \mid 1\leq j<k}\subseteq \dom(T)\). {A (possibly infinite) sequence \(\pi\) over $\N$ is a \emph{path} of \(T\) just if every finite prefix of \(\pi\) is in \(\dom(T)\).} A tree is \emph{ranked} just if \(\max \, \set{j \; | \; s
  \, j\in \dom(T)}\) is equal to the arity of \(T(s)\) for each \(s\in
\dom(T)\).



\subsubsection*{Higher-Order Recursion Schemes}
The set of \emph{\kinds{}} is defined by:
\iffull
\[
\kappa ::= \T \mid \kappa_1 \ra \kappa_2
\]
\fi
where \(\T\) is the type of trees. {By convention, $\ra$ associates to the right; thus, for example, $o \ra o \ra o$ means $o \ra (o \ra o)$.} The \emph{order} of \(\kappa\), written \(\order(\kappa)\), is defined by:
\iffull
\[
\begin{array}{rll}
\order(\T) & := & 0 \\
\order(\kappa_1\ra\kappa_2) & := & \max \, (\order(\kappa_1)+1, \order(\kappa_2)).\\
\end{array}
\]
\fi
A (deterministic) \emph{higher-order recursion scheme} (recursion scheme, for short)
is a {quadruple} \(\GRAM = (\TERMS, \NONTERMS, \RULES, S)\), where

\begin{enumerate}[(i)]
\item \(\TERMS\) is a {ranked alphabet} {of \emph{terminal symbols}}. 
\item \(\NONTERMS\) is a {map} from a finite set of symbols called \emph{non-terminals} to \kinds{}.
\item \(\RULES\) is a set of rewrite rules \(F\;\seq{x}\Hra t\). Here {\(\seq{x} = x_1, \cdots, x_n\)} abbreviates a sequence of variables, and \(t\) is {an applicative} term constructed from non-terminals, terminals, and variables $x_1, \cdots, x_n$. 
\item \(S\) is a \emph{start symbol}.
\end{enumerate}
We require that \(\NONTERMS(S) = \T\).
The set of (typed) terms is defined in the standard manner:
A non-terminal or variable of \kind{} \(\kappa\) is a term of \kind{} \(\kappa\).
A terminal of arity \(k\) is a term of \kind{} \(\underbrace{\T\ra \cdots \ra \T}_k\ra \T\).
If terms \(t_1\) and \(t_2\) have \kinds{} \(\kappa_1\ra\kappa_2\) and \(\kappa_1\) respectively,
then \(t_1\ t_2\) is a term of \kind{} \(\kappa_2\). 
{By convention, application associates to the left; thus, for example, $s \, t \, u$ means $(s \, t) \, u$.} For each rule \(F\;\seq{x}\Hra t\),
\(F\ \seq{x}\) and \(t\) must be terms of \kind{} \(\T\).
There must be exactly one rewrite rule for each non-terminal.
The \emph{order} of a recursion scheme is the highest order of (the types of) its non-terminals.

A rewrite relation on terms is defined inductively by:
\iffull
\begin{enumerate}[(i)]
\else
\begin{inparaenum}[(i)]
\fi
\item If $F\,\seq{x}\Hra t\in \RULES$, then \(F \, \seq{s} \Hred{\GRAM} \subst{\seq{x}}{\seq{s}}t\).
\item If \(t\Hred{\GRAM} t'\), then \(t\,s\Hred{\GRAM}t'\,s\) and \(s\,t\Hred{\GRAM}s\,t'\).
\iffull
\end{enumerate}
\else
\end{inparaenum}
\fi
The \emph{value tree} of a recursion scheme \(\GRAM\), written \(\sem{\GRAM}\),
is the (possibly infinite) tree obtained by infinite rewriting of the start symbol \(S\).
More precisely, let us define \(\Bot{t}\) by:
\[
 \Bot{a} := a \qquad \Bot{F} := \bot \qquad
\Bot{(t_1 t_2)} := \left\{\begin{array}{ll}
     \bot & \mbox{if \(\Bot{t_1}=\bot\)}\\
     \Bot{t_1}\,\Bot{t_2} & \mbox{otherwise}
  \end{array}\right.
\]
The value tree \(\sem{\GRAM}\) is the \(\Sigma\cup\set{\bot\mapsto 0}\)-ranked tree defined by:
\[ \sem{\GRAM} \; := \; \bigsqcup \set{\Bot{t} \mid S \Hreds{\GRAM} t}.\]
Here, \(\bigsqcup S\) denotes the least upper bound with respect to the tree order \(\sqsubseteq\) defined by
\[ T_1 \sqsubseteq T_2 \; \IFF \; \forall s\in\dom(T_1) \, . \, (T_1(s)=T_2(s) \; \lor \; T_1(s)=\bot)\]
Note that \(\sem{\GRAM}\) is always well-defined, as the rewrite relation \(\Hred{\GRAM}\) is confluent.

\begin{example}
\label{ex:recursion-scheme}
Consider the recursion scheme \(\GRAM = (\TERMS,\NONTERMS,\RULES,S)\) where
\[
\begin{array}{l}
\TERMS = \set{\Ta\mapsto 2, \ \Tb\mapsto 1, \ \Tc\mapsto 1,\ \Te\mapsto 0}\\
\NONTERMS = \set{S\mapsto \T, F\mapsto (\T\ra\T)\ra \T\ra \T, I\mapsto \T\ra\T, C\mapsto (\T\ra\T)\ra(\T\ra\T)\ra(\T\ra\T)}\\
\RULES = \{\\
\quad S \Hra F\, I\, \Te,\\
\quad F\,f\,x \Hra \Ta\, (f\, x)\, (F\,(C\, \Tb\,f)\, (\Tc\,x)),\\
\quad I\,x \Hra x,\\
\quad C\,f\,g\,x \Hra f(g\, x)\\
\}
\end{array}
\]
\(S\) is reduced as follows.
\[
\begin{array}{rll}
S & \red{} & F\,I\,\Te \\
& \red{} & \Ta\, (I\,\Te)\, (F\,(C\,\Tb\,I)\, (\Tc\,\Te))\\
&   \red{} & \Ta\, \Te\, (\Ta\,(C\,\Tb\,I\,(\Tc\,\Te))\, (F\,(C\,\Tb\,(C\,\Tb\,I)))\, (\Tc\,(\Tc\,\Te)))\\
& \reds{} & \Ta\, \Te\, (\Ta\,(\Tb\,(\Tc\,\Te))\, (F\,(C\,\Tb\,(C\,\Tb\,I)))\, (\Tc\,(\Tc\,\Te)))\\
  & \reds{} & \Ta\,\Te\, (\Ta\,(\Tb\,(\Tc\,\Te)) (\Ta\, (\Tb^2 (\Tc^2\,\Te)) (\Ta\, (\Tb^3 (\Tc^3\,\Te))\, \cdots)))\\
\end{array}
\]
The value tree is shown in Figure~\ref{fig:valtree}.
Each path of the tree is labelled by \(\Ta^{m+1}\Tb^{m}\Tc^{m}\Te\).
\end{example}
\begin{figure}
\[
\Tree[ [ ].{$\Te$} [ [ [ [ ].{$\Te$} ].{$\Tc$} ].{$\Tb$} [ [ [ [ [ [ ].{$\Te$}  ].{$\Tc$}  ].{$\Tc$} ].{$\Tb$} ].{$\Tb$} [ [ [ [ [ [ ].{$\cdots$}  ].{$\Tc$}  ].{$\Tb$} ].{$\Tb$} ].{$\Tb$} [ [ [ [ [ ].{$\cdots$}  ].{$\Tb$}  ].{$\Tb$} ].{$\Tb$} [ [ [ [ ].{$\cdots$}  ].{$\Tb$} ].{$\Tb$} [ [ [ ].{$\cdots$}  ].{$\Tb$} [ ].{$\cdots$} ].{$\Ta$}  ].{$\Ta$} ].{$\Ta$} ].{$\Ta$}  ].{$\Ta$} ].{$\Ta$} ].{$\Ta$}
\]
\label{fig:valtree}
\caption{The tree generated by the recursion scheme of Example~\ref{ex:recursion-scheme}}
\end{figure}

\subsubsection*{Alternating parity tree automata}
Given a finite set $X$, the set ${\sf B}^+(X)$ of \emph{positive Boolean formulas} over $X$ is defined as follows. We let $\theta$ range over ${\sf B}^+(X)$.
\[\theta \; ::= \; {\sf t} \; | \; {\sf
f} \; | \; x \; | \; \theta \wedge \theta \; | \; \theta \vee
\theta \]
{{where} \(x\) ranges over \(X\).}
We say that a subset $Y$ of $X$ \emph{satisfies} $\theta$ just if
assigning true to elements in $Y$ and false to elements in $X
\setminus Y$ makes $\theta$ true. 

An \emph{alternating parity tree automaton} (or APT for short) over $\Sigma$-labelled trees is a tuple ${\cal A} \; = \; (\Sigma, Q, \delta, q_I, \Omega)$ where
\iffull
\begin{enumerate}[(i)]
\else
\begin{enumerate}[(i)]
\fi
\item $\Sigma$ is a {ranked alphabet}; let $m$ be the largest arity of the terminal symbols;
\item $Q$ is a finite set of states, and $q_I \in Q$ is the initial state;
\item $\delta : Q \times \Sigma \longrightarrow {\sf B}^+(\set{1, \cdots, m}\times Q)$ is the transition function where, for each $f \in \Sigma$ and $q \in Q$, we have $\delta(q, f) \in {\sf B}^+(\set{1, \cdots, \arity(f)} \times Q)$; and
\item $\Omega : Q \longrightarrow {\set{0, \cdots, M-1}}$ is the priority function.
\iffull
\end{enumerate}
\else
\end{enumerate}
\fi



A \emph{run-tree} of an APT $\cal A$ over a $\Sigma$-labelled ranked tree $T$ is a $(\dom(T) \times Q)$-labelled unranked tree $r$ satisfying:
\iffull
\begin{enumerate}[(i)]
\else
\begin{inparaenum}[(i)]
\fi
\item $\epsilon \in \dom(r)$ and $r(\epsilon) = (\epsilon, q_I)$; and
\item for every $\beta \in \dom(r)$ with $r(\beta) = (\alpha, q)$, there is a set $S$ that satisfies $\delta(q, {T}(\alpha))$;
and for each $({i}, q') \in S$, there is some ${j}$ such that $\beta \, {j} \in \dom(r)$
and $r(\beta \, {j}) = (\alpha \, {i}, q')$.
\iffull
\end{enumerate}
\else
\end{inparaenum}
\fi

Let $\pi = \pi_1 \, \pi_2 \, \cdots$ be an infinite path in $r$; for each $i \geq 0$, let the state label of the node $\pi_1 \cdots \pi_i$ be $q_{n_i}$ {where $q_{n_0}$, the state label of $\epsilon$, is $q_I$}. We say that $\pi$ satisfies the \emph{parity} condition just if the largest priority that occurs infinitely often in $\Pfun(q_{n_0}) \, \Pfun(q_{n_1}) \, \Pfun(q_{n_2}) \cdots$ is even. A run-tree $r$ is \emph{accepting} if every infinite path in it satisfies the parity condition.
An APT \(\APT\) accepts a (possibly infinite) ranked tree \(T\) if there is an accepting run-tree of \(\APT\) over \(T\).

Ong~\cite{Ong06LICS} has shown that there is a procedure that, given a recursion scheme \(\GRAM\) and an APT \(\APT\), decides whether $\APT$ accepts the value tree of \(\GRAM\).

\begin{thm}[Ong]
\label{th:Ong}
Let \(\GRAM\) be a recursion scheme of order \(n\), and \(\APT\) be an APT. The problem of deciding whether $\APT$ accepts \(\sem{\GRAM}\) is \(n\)-EXPTIME-complete.
\end{thm}



As usual (following \cite{Ong06LICS}),
we restrict our attentions to recursion schemes whose value trees do not contain \(\bot\)
in the rest of the paper. Given a recursion scheme \(\GRAM\) that may generate \(\bot\) and an APT \(\APT\),
one can construct \(\GRAM'\) and \(\APT'\) such that (i) \(\APT\) accepts \(\sem{\GRAM}\) if and only if \(\APT'\) accepts \(\sem{\GRAM'}\),
and (ii) \(\GRAM'\) does not generate \(\bot\). \footnote{Note, however, that the transformation does not preserve the class of trivial APT considered in Section~\ref{sec:ta-hardness}.}



%% file: trivialAPT.tex
\section{Trivial APT and the Complexity of Model Checking}
\label{sec:ta-hardness}


\emph{APT with a trivial acceptance condition}, or \emph{trivial APT}
(for short), is an APT that has exactly one priority which is
even. Note that trivial APT are equivalent to Aehlig's
``trivial automata''~\cite{Aehlig07} (for defining languages of ranked
trees).


\newcommand\nuform[2]{\nu{#1}.{#2}}
\newcommand\muform[2]{\mu{#1}.{#2}}
\newcommand\sform[3]{{#1}{#2}.{#3}}

\newcommand\diaform[2]{{\langle #1 \rangle}#2}
\newcommand\subform[1]{\mathit{SubF}(#1)}
\newcommand\anglebra[1]{\langle\,#1\rangle}
\newcommand\powerset[1]{{\mathcal P}(#1)}
\newcommand{\Eloise}{\'Elo\"ise\xspace}
\newcommand{\Abelard}{Abelard\xspace}
\newcommand\encode[1]{{\ulcorner {#1} \urcorner}}

The first result of this paper is a logical characterization of the class of {$\Sigma$-labelled} ranked trees accepted by
trivial APT. Call $\cal S$ the following fragment of the modal mu-calculus:
\[\phi, \psi \; ::= \; \TRUE 
\; | \; \FALSE
\; | \; P_f
\; | \; Z
\; | \; \phi \wedge \psi
\; | \; \phi \vee \psi
\; | \; \diaform{i}{\phi}
\; | \; \nuform{Z}{\phi}
\]
where $f$ ranges over symbols in a $\Sigma$, and $i$ ranges over $\set{1, \cdots, \arity(\Sigma)}$. (We think of $\cal S$ as the ``safety'' fragment.)
\iffull We give a characterization of trivial APT. 
 A proof is given in Appendix~\ref{sec:equi-expressivity}. \fi


\begin{prop}[Equi-Expressivity]
\label{th:equi-expressivity}
The logic $\cal S$ and 
trivial APT are equivalent for defining possibly-infinite
ranked trees. I.e.~for every closed $\cal S$-formula, there is a trivial APT that defines the
same tree language, and vice versa.
\end{prop}


We show that the model checking problem for recursion schemes is \(n\)-EXPTIME complete
for trivial APT.
The upper-bound of \(n\)-EXPTIME follows immediately from Ong's result~\cite{Ong06LICS}.
To show the lower-bound,
we reduce the decision problem of \(w\stackrel{?}{\in} \lang(\APDA)\), where \(w\) is a word
and \(\APDA\) is an order-\(n\) alternating PDA, to the model checking problem for recursion schemes.
\(n\)-EXPTIME hardness follows from the reduction, since
the problem of \(w\stackrel{?}{\in} \lang(\APDA)\)
is
\(n\)-EXPTIME hard~\cite{Engelfriet91}.



\begin{defi}\label{def:apda}
An \emph{order-\(n\) alternating PDA} (order-\(n\) APDA, for short) for {finite} words is a 7-tuple:
\[\APDA \; = \; \langle P, \; \lambda, \; p_0, \; \Gamma, \; {\Sigma}, \; \Delta, \; F\rangle\]
where \(P\) is a set of states, {\(\lambda : P \ra \set{\UNIV,\EXISTS}\) partitions states into universal and existential}, 
\(p_0\) is the initial state, \(\Gamma\) is a stack alphabet, \(\Sigma\) is an input alphabet,
\(F \subseteq P\) is the set of final states,
and \(\Delta  \subseteq P\times \Gamma\times ({\Sigma}\cup\set{\epsilon})\times P\times \OP{n}\) is a transition relation.
A \emph{configuration} of an order-\(n\) APDA is of the form
\((p, s)\) where \(s\) is an order-\(n\) stack (an order-\(1\) stack is an ordinary stack, and
an order-\((k+1)\) stack is a stack of order-$k$ stacks).
The induced transition relation on configurations is defined by the rule:
\[ \mbox{if \((p, \topn{1}(s), \alpha, p', \theta)\in \Delta\),
then } (p,\stack) \red{\alpha} (p', \theta(\stack))\]
where \(\theta\in \OP{n}\) is an order-$n$ stack operation%
\footnote{Assume an order-$n$ stack, where $n \geq 2$. An
  \emph{order-1 push} operation is just the standard operation that pushes a
  symbol onto the top of the top order-1 stack; the \emph{order-1 pop} operation removes the
  top symbol from the top order-1 stack. For $2 \leq i \leq n$, the \emph{order-$i$ push} operation
  duplicates the top order-$(i-1)$ stack of the order-$n$ stack;
  the \emph{order-$i$ pop} operation removes the top order-$(i-1)$ stack. The
  set $\OP{n}$ of order-$n$ stack operations consists of order-$i$
  push and order-$i$ pop for each $1 \leq i \leq n$. For a formal
  definition, see, for example, the FoSSaCS 2002 paper~\cite{Knapik02FOSSACS} of Knapik et al.}
and \(\topn{1}(s)\) is the stack top of \(s\).

Let \(w\) be a word over \(\Sigma\). We write \(w_i\) ({where} \(0\leq i< |w|\)) for the \(i\)-th element of \(w\).
A \emph{run tree}
of an order-$n$ APDA over a word \(w\) is
a \emph{finite}, unranked tree {satisfying the following}.
\begin{enumerate}[(i)]
\item The root is labelled by \((p_0, \bot_n, 0)\), where \(\bot_n\) is the {empty order-$n$ stack}.
\item If a node is labelled by \((p,s,i)\), then one of the following conditions holds,
where 
\[\eqalign{
  \Xi 
:={}& \set{(p',\theta(s),i+1)\mid (p,\stacktop(s),w_i,p',\theta)\in
    \Delta \land i<|w|}\cr
&   \cup \set{(p',\theta(s),i)\mid
    (p,\stacktop(s),\epsilon,p',\theta)\in \Delta}.
}\]
\begin{iteMize}{$\bullet$}
\item \(p\in F\) and \(i=|w|\);
\item \(\lambda(p)=\UNIV\) and the set of labels of the child nodes is $\Xi$; or
\item \(\lambda(p)=\EXISTS\) and there is exactly one child node, which is labelled by an element of $\Xi$.
\end{iteMize}
 (It follows that the leaves of a run tree are labelled by
$(p, s, |w|)$ with $p \in F$, or $(p, s, i)$ with $\lambda(p) = {\tt A}$ and $\Xi = \emptyset$.) 
\end{enumerate}
An order-$n$ APDA \(\APDA\) \emph{accepts} \(w\) if there exists a run tree of \(\APDA\) over \(w\).%
\end{defi}

Engelfriet \cite{Engelfriet91} has shown that the word acceptance problem for order-$n$ APDA is \(n\)-EXPTIME complete.
\begin{thm}[Engelfriet]
\label{th:nexptime-of-APDA}
Let \(\APDA\) be an order-\(n\) APDA and \(w\) a finite word over \(\Sigma\).
The problem of \(w\stackrel{?}{\in}\lang(\APDA)\) is \(n\)-EXPTIME complete.
\end{thm}

To reduce the word acceptance problem of order-\(n\) APDA to the model
checking problem for recursion schemes, we use the equivalence
\cite{Knapik02FOSSACS} between order-$n$ \emph{safe}%
\footnote{An order-\(n\) recursion scheme is \emph{safe} if it satisfies 
a certain condition called safety~\cite{KNU02}. We use the equivalence between safe recursion
schemes and higher-order PDA just to prove the lower-bound, so that the knowledge about
the safety constraint is not required. See \cite{KNU02,BO07} for details of the safety constraint.
\nk{I shortened the footnote. Is this OK?} \lo{Yes it is.}}
recursion schemes and order-\(n\) PDA as (deterministic) devices for generating trees.

\begin{defi}\label{def:treecpda}
{An \emph{order-$n$ tree-generating PDA} is a tuple $\mathcal{A} = \anglebra{\Sigma, \Gamma, Q, \delta, q_0}$ where $\Sigma$ is a ranked alphabet, $\Gamma$ is a stack alphabet, $Q$ is a finite set of states, $q_0 \in Q$ is the initial state, and
    \[\delta \, : \, Q \times \Gamma \,
  \longrightarrow \, (Q \times \mathit{Op}_n \; \cup \; \set{(f ;
    q_1, \cdots, q_{\arity(f)}) \mid f \in \Sigma, q_i \in Q})\]
is the transition function. A \emph{configuration} is either a pair $(q, s)$ where $q \in Q$ and $s$ is an order-$n$ stack, or a triple of the form $(f; q_1 \cdots q_{\arity(f)}; s)$ where $f \in \Sigma$ and $q_1 \cdots q_{\arity(f)} \in Q^\ast$. Let $\overline \Sigma$ be the label-set $\set{(f, i) \mid f \in \Sigma, 1 \leq i \leq \arity(f)} \cup \set{a \in \Sigma \mid \arity(a) = 0}$. We define the labelled transition relation between configurations induced by $\delta$:
\renewcommand\arraystretch{1.3}
\[\begin{array}{l}
(q, s) \redlab{\epsilon} (q', \theta(s)) \quad
\hbox{if $\delta(q, \topn{1}(s)) = (q', \theta)$}\\
(q, s) \redlab{\epsilon} (f; \overline q; s) \quad
\hbox{if $\delta(q, \topn{1}(s)) = (f; \overline q)$ and $\arity(f) \geq 1$}\\
(q, s) \redlab{a} (a; \epsilon; s) \quad
\hbox{if $\delta(q, \topn{1}(s)) = (a; \epsilon)$ and $\arity(a) = 0$}\\
(f; \overline q; s) \redlab{(f, i)} (q_i, s) \quad
\hbox{where $1 \leq i \leq \arity(f)$}\\
\end{array}\]\medskip

\noindent Let $w$ be a finite or infinite word over the alphabet $\overline \Sigma$. We say that $w$ is a \emph{trace} of $\mathcal{A}$ just if there is a possibly-infinite sequence of transitions $(q_0, \bot_n) \redlab{\ell_1} \gamma_1 \cdots \redlab{\ell_m} \gamma_m \redlab{\ell_{m+1}} \cdots$ such that $w = \ell_1 \ell_2 \cdots$. We say that $\cal A$ \emph{generates} a $\Sigma$-labelled tree $t$ just in case
the {branch language}\footnote{The \emph{branch language} of
  $t : \dom(t) \longrightarrow \Sigma$ consists of
  \begin{enumerate}[(i)]
  \item infinite words $(f_1,
  d_1)(f_2, d_2) \cdots $ just if there exists $d_1 \, d_2 \cdots \in \set{1, 2, \cdots, m}^\omega$ (where $m$ is the maximum arity of the $\Sigma$-symbols) such that $t(d_1
  \cdots d_i) = f_{i+1}$ for every $i \geq 0$; and
  \item finite words $(f_1, d_1) \cdots (f_n,
  d_n) \, f_{n+1}$ just
if there exists $d_1 \cdots d_n \in \set{1, \cdots, m}^\ast$ such that $t(d_1 \cdots d_i) =
  f_{i+1}$ for $0 \leq i \leq n$, and the arity of \(f_{n+1}\) is \(0\).
  \end{enumerate}}  of $t$ coincides with the set of \changed{maximal} traces of $\cal A$.}
\end{defi}

\begin{thm}[Knapik et al.~\cite{Knapik02FOSSACS}]
\label{th:PDA-and-recursion-scheme}
There is an {effective transformation that, given an order-\(n\) tree-generating PDA \(\M\), returns an order-\(n\) safe recursion scheme
\(\GRAM\)} that generates the same tree as \(\M\).
Moreover, both the running time of the transformation algorithm and the size of \(\GRAM\) are polynomial in the size of \(\M\).
\end{thm}

By Theorems~\ref{th:nexptime-of-APDA} and \ref{th:PDA-and-recursion-scheme},
it suffices to show that, given a word \(w\) and an order-\(n\) APDA \(\APDA\), one can construct
an order-\(n\) tree-generating PDA \(\trPDA{\APDA}{w}\) and a trivial APT \(\trBUCHI{\APDA}{}\)
such that \(w\) is accepted by \(\APDA\) if, and only if,
the tree generated by \(\trPDA{\APDA}{w}\) is accepted by \(\trBUCHI{\APDA}{}\).

Let \(w\) be a word over \(\Sigma\).
From \(w\) and \(\APDA = \anglebra{P, \lambda, p_0, \Gamma, {\Sigma}, \Delta, F}\) above,
we construct an order-\(k\) PDA \(\trPDA{\APDA}{w}\) for generating a $\set{\UNIV, \EXISTS, \BOT, \TOP}$-labelled tree, which is a kind of
run tree of \(\APDA\) over the input word \(w\).
The node label \(\UNIV\) (\(\EXISTS\), respectively) means
that \(\APDA\) is in a universal (existential, respectively) state; \(\TOP\) means that
\(\APDA\) has accepted the word, and \(\BOT\) means that \(\APDA\) is stuck (no outgoing transition).

Let \(N\) \changed{be}
\(\MAX_{q\in P, a\in \Sigma, \gamma\in \Gamma} |
 \set{(q',a',\theta) \mid (q, \gamma, a', q', \theta)\in\Delta, a'\in \set{a, \epsilon}}|\).
{I.e.}~\(N\) is the degree of non-determinacy of \(\APDA\).
We define \[ \trPDA{\APDA}{w} = \langle {\set{\UNIV \mapsto N, \EXISTS \mapsto N, \TOP \mapsto 0, \BOT \mapsto 0}}, \Gamma, Q, \delta, (p_0, 0)\rangle\]
where:
\begin{iteMize}{$-$}
\item \(Q = (P \times \set{0,\ldots,{|w|}})\; \cup \; \set{q_\top, q_\bot} \; \cup \;
   (P \times \set{0,\ldots,{|w|}}\times \OP{n})\)
\item \(\delta : Q \times \Gamma \longrightarrow (Q \times \mathit{Op}_n \changed{\cup} 
\changed{\set{(g ; q_1,\ldots,q_k) :
g \in \set{\UNIV, \EXISTS, \TOP, \BOT}, k\geq 0, q_1,\ldots,q_k \in Q}})\) is given by:
\[
\begin{array}{ll}
{(1)}\quad\delta((p,|w|), \gamma) = {(\TOP; \epsilon)}, \mbox{ if $p\in F$}
   \\ 
(2)\quad\delta((p, i), \gamma) = (\UNIV; (p_1,j_1, \theta_1),\ldots, (p_m, j_m, \theta_m), \underbrace{q_\top, \ldots, q_\top}_{N-m})\\
\qquad    \mbox{if } \lambda(p)=\UNIV \mbox{ and }
\set{(p_1,j_1, \theta_1),\ldots,(p_m,j_m, \theta_m)} \mbox{ is: }\\
\qquad   \set{(p',i+1,\theta) \mid (p, \gamma, w_i, p',\theta)\in \Delta\land i<|w|}
    \cup \set{(p',i,\theta) \mid (p, \gamma, \epsilon, p',\theta)\in \Delta}\\
(3)\quad\delta((p, i), \gamma) = (\EXISTS; (p_1,j_1, \theta_1),\ldots, (p_m, j_m, \theta_m), {\underbrace{q_\bot, \ldots, q_\bot}_{N-m}})\\
\qquad    \mbox{if }\lambda(p)=\EXISTS \mbox{ and }
 \set{(p_1, j_1, \theta_1),\ldots,(p_m,j_m, \theta_m)} \mbox{ is: }\\
\qquad  \set{(p',i+1,\theta) \mid (p, \gamma, w_i, p',\theta)\in \Delta\land i<|w|}
  \cup \set{(p',i, \theta) \mid (p, \gamma, \epsilon, p',\theta)\in \Delta}\\
(4)\quad\delta((p,i,\theta), \gamma) = ((p,i), \theta)\\
(5)\quad\delta(q_\top, \gamma) = (\TOP; {\epsilon})\\
(6)\quad\delta(q_\bot, \gamma) = (\BOT; {\epsilon})\\
\end{array}
\]
\end{iteMize}\medskip


\noindent Rules {(2)} and {(3)} are applied only when rule (1) is inapplicable.
\(\trPDA{\APDA}{w}\) simulates \(\APDA\) over the word \(w\), and constructs a tree representing
the computation of \(\APDA\). A state \((p, i)\in P\times \set{0,\ldots,|w|-1}\) simulates
 \(\APDA\) in state \(p\) reading the letter \(w_i\).
A state \((p, i, \theta)\) simulates an intermediate transition state of \(\APDA\), where $\theta$ is the stack operation to be applied.
The states \(q_\top\) and \(q_\bot\) are for creating dummy subtrees of nodes labelled with \(\UNIV\) or \(\EXISTS\), so that the number of children of these nodes adds up to $N$, the arity of \(\UNIV\) and \(\EXISTS\).
Rule (1) ensures that
when \(\APDA\) has read the input word and reached a final state,
\(\trPDA{\APDA}{w}\) stops simulating \(\APDA\) and outputs \(\TOP\).
Rule (2) is used to simulate
transitions of \(\APDA\) in a universal state, reading the \(i\)-th input:
\(\trPDA{\APDA}{w}\) constructs a node labelled \(\UNIV\) (to record that \(\APDA\)
was in a universal state) and spawns threads to simulate all possible
transitions of \(\APDA\).
Rule (3) is for simulating \(\APDA\) in an existential state.
Note that, if \(\APDA\) gets stuck (i.e.~if there is no outgoing transition),
all children of the \(\EXISTS\)-node are labelled \(\BOT\); thus failure of
the computation can be recognized by the trivial APT given in the following.
Rule (4) is just for intermediate transitions. Note that
 a transition of \(\APDA\) is simulated by \(\trPDA{\APDA}{w}\)
in two steps: the first for {outputting} \(\UNIV\) or \(\EXISTS\), and
the second for changing the stack.

Now we construct a trivial APT \(\trBUCHI{\APDA}{}\) that accepts the tree generated by \(\trPDA{\APDA}{w}\) if, and only if,
\(w\) is \emph{not} accepted by \(\APDA\).
\iffull
The trivial APT \(\trBUCHI{\APDA}{}\) is given by:
\[
\trBUCHI{\APDA}{} := \anglebra{ \set{\UNIV, \EXISTS, \TOP, \BOT}, \set{q_0}, \changed{\delta, q_0}, \set{q_0\mapsto 0}}
\]
where:
\fi
\[
\begin{array}{l}
\delta(q_0, \UNIV) = \bigvee_{i=1}^{N}(i, q_0) \quad  
\delta(q_0, \EXISTS) = \bigwedge_{i=1}^{N}(i, q_0)\quad
\delta(q_0, \TOP) = {{\sf f}} \quad
\delta(q_0, \BOT) = {\TRUE}
\end{array}
\]
Intuitively, \(\trBUCHI{\APDA}{}\) accepts all trees representing a failure computation tree of \(\APDA\).
If the automaton in state \(q_0\) reads \(\TOP\) (which corresponds to an accepting state of
\(\APDA\)), it gets stuck. Upon reading \(\UNIV\), the automaton non-deterministically chooses one of the
subtrees, and checks whether the subtree represents a failure computation of \(\APDA\).
On the other hand, upon reading \(\EXISTS\), the automaton checks that all subtrees represent
failure computation trees of \(\APDA\).

{By the above construction, we have:}
\begin{thm}
Let \(w\) be a word, and \(\APDA\) an order-\(n\) APDA.
{Then \(w\) is \emph{not} accepted by \(\APDA\) if, and only if,
the tree generated by \(\trPDA{\APDA}{w}\) is accepted by \(\trBUCHI{\APDA}{}\).}
\end{thm}

\begin{cor}
\label{cor:hardness-trivial-APT}
The trivial APT acceptance problem for
  the tree generated by an order-$n$ recursion scheme (i.e.~whether the
  tree generated by a given order-$n$ recursion scheme is accepted by
  a given trivial APT) is $n$-EXPTIME hard in the size of the
  recursion scheme.
\end{cor}


By modifying the encoding, we can also show that the model checking problem is
\(n\)-EXPTIME-hard in the size of the APT.
The idea is to modify \(\trPDA{\APDA}{w}\) so that it generates a tree representing
computation of \(\APDA\) over not just \(w\) but all possible input words, and let
a trivial APT check the part of the tree corresponding to the input word \(w\).
As a result, the trivial APT depends on the input word \(w\), but the tree-generating PDA does not.

We make the following two assumptions on \(\APDA\) (without loss of generality):
\begin{enumerate}[(i)]
\item In each state, if \(\APDA\) can perform an \(\epsilon\)-transition, then
\(\APDA\) cannot perform any input transition
i.e.~\(\set{(p',\theta)\mid \exists a\in\Sigma.(p,\gamma,a,p',\theta)\in \Delta}\neq \emptyset\)
implies
\(\set{(p',\theta)\mid (p,\gamma,\epsilon,p',\theta)\in \Delta}= \emptyset\).

\item {There is no transition from a final state
i.e.~if $p \in F$ then \(\set{(p',\theta)\mid \exists a\in\Sigma\cup\set{\epsilon}.(p,\gamma,a,p',\theta)\in \Delta}= \emptyset\).}
\end{enumerate}\medskip

\noindent Given an order-$n$ APDA \(\APDA\) and a word \(w\),
we shall construct \(\tPDA'_{\APDA}\) and \(\trBUCHI{}{w}\), such that
\(w\) is \emph{not} accepted by \(\APDA\) if, and only if,
the tree generated by {\(\tPDA'_{\APDA}{}\)} is accepted by \(\trBUCHI{}{w}\).
The difference from the construction of \(\trPDA{\APDA}{w}\) and \(\trBUCHI{\APDA}{}\) above is
that  \(\tPDA_{\APDA}'\) {does} not depend on \(w\).
The idea is to let \(\tPDA_{\APDA}'\) generate a tree representing
the computations of \(\APDA\) over all possible inputs. We then let \(\trBUCHI{}{w}\)
traverse the part of the tree corresponding to the computation over \(w\), and check
whether the computation is successful.

We define a tree-generating PDA
\( \tPDA_{\APDA}' = \langle \Sigma', \Gamma, Q, \delta, q_0\rangle\) where:
\begin{itemize}[-]
\item \(\Sigma' = {\set{\READ \mapsto |\Sigma|,\ACCEPT \mapsto 0,\EPSILON \mapsto 1, \UNIV \mapsto N, \EXISTS \mapsto N, \TOP \mapsto 0, \BOT \mapsto 0}}\) 
\item \(Q = P \; \cup\; (P \times (\Sigma\cup\set{\epsilon}))\; \cup \; \set{q_\top, q_\bot} \; \cup \;
  (P \times \OP{k})\)
\item \(q_0 = p_0\)
\item \(\delta\) is given by:
\[
\begin{array}{ll}
\delta(p, \gamma) = (\ACCEPT; \epsilon) \mbox{ if $p\in F$}\\
\delta(p, \gamma) = (\EPSILON; ((p, \epsilon), \ID))
   \mbox{ if $\set{(p',\theta)\mid (p,\gamma,\epsilon,p',\theta)\in \Delta}\neq \emptyset$}.\\
\delta(p, \gamma) = (\READ; ((p, a_1), \ID),\ldots,((p,a_n),\ID))\\
\qquad   \mbox{ if $p\not\in F$,
$\set{(p',\theta)\mid (p,\gamma,\epsilon,p',\theta)\in \Delta}= \emptyset$
and  $\Sigma = \set{a_1,\ldots,a_n}$}.\\
\delta((p, \alpha), \gamma) = (\UNIV; ((p_1, \theta_1),\ldots, (p_m, \theta_m), q_\top, \ldots, q_\top))\\
\qquad    \mbox{if }\lambda(p)=\UNIV \mbox{ and}\\
\qquad\quad \set{(p_1, \theta_1),\ldots,(p_m,\theta_m)} =
    \set{(p',\theta) \mid (p, \gamma, \alpha, p',\theta)\in \Delta}\\
\delta((p, \alpha), \gamma) = (\EXISTS; ((p_1,\theta_1),\ldots, (p_m, \theta_m), q_\bot, \ldots, q_\bot))\\
\qquad    \mbox{if }\lambda(p)=\EXISTS \mbox{ and}\\
\qquad\quad \set{(p_1, \theta_1),\ldots,(p_m,\theta_m)} =
    \set{(p',\theta) \mid (p, \gamma, \alpha, p',\theta)\in \Delta}\\
\delta((p,\theta), \gamma) = (p, \theta)\\
\delta(q_\top, \gamma) = (\TOP; {\epsilon})\\
\delta(q_\bot, \gamma) = (\BOT; {\epsilon})\\
\end{array}
\]
\end{itemize}
In a final state of \(\APDA\), \(\tPDA_{\APDA}'\) outputs a {node} labelled with \(\ACCEPT\),
to indicate that \(\APDA\) has reached a final state, and stops simulating \(\APDA\) (as,
by assumption (ii) above, there is no outgoing transition).
In a state where \(\APDA\) has \(\epsilon\)-transitions,
 \(\tPDA_{\APDA}'\) outputs a node labelled with \(\EPSILON\), and then simulates all the possible \(\epsilon\)-transitions of \(\APDA\).
In a state where \(\APDA\) has input transitions,
 \(\tPDA_{\APDA}'\) outputs a node labelled with \(\READ\) to indicate that \(\APDA\) makes
an input transition, and then simulates the input transition for each possible input symbol.
Note that by the assumptions 
(i) and (ii) above, these three transitions are disjoint.
The remaining transition rules are analogous to those of \(\trPDA{\APDA}{w}\).

Define the trivial APT \(\trBUCHI{}{w}\) by
$\trBUCHI{}{w} = \anglebra{\Sigma', Q', \delta, q_0, \Pfun}$ where:
\[
\begin{array}{l}
Q' = \set{q_0, \ldots, q_{|w|}}\\
\delta(q, \EPSILON) = (1, q) \mbox{ for every $q\in Q'$}\\
\delta(q_i, \READ) = (j, q_{i+1}) \mbox{ if $0\leq i\leq |w|-1$ and $w_i = a_j$}\\
\delta(q_{|w|}, \READ) = \TRUE \\
\delta(q_i, \UNIV) = (1, q_i)\lor \cdots \lor (N, q_i)\\
\delta(q_i, \EXISTS) = (1, q_i)\land \cdots \land (N, q_i)\\
\delta(q_{|w|}, \ACCEPT) = \FALSE\\
\delta(q_{i}, \ACCEPT) = \TRUE \mbox{ for every $0\leq i< |w|$}\\
\delta(q, \TOP) = \FALSE \mbox{ for every $q\in Q'$}\\
\delta(q, \BOT) = \TRUE \mbox{ for every $q\in Q'$}\\
\end{array}
\]
and \(\Pfun\) is the trivial priority function.

The trivial APT \(\trBUCHI{}{w}\) traverses the tree generated by \(\tPDA_{\APDA}'\) (which represents
transitions of \(\APDA\) for all possible inputs), while keeping track of the position of the input head of \(\APDA\) in its state
(\(q_i\) means that \(\APDA\) is reading the \(i\)-th letter of the word \(w\)).
Upon reading \(\READ\) in state \(q_i\), \(\trBUCHI{}{w}\) proceeds to traverse the branch corresponding to
the \(i\)-th letter (i.e.~\(w_i\)).
Reading \(\ACCEPT\) in state \(q_{|w|}\) means that \(\APDA\) accepts the word \(w\), so that
the run of \(\trBUCHI{}{w}\) fails (recall that \(\trBUCHI{}{w}\) accepts the tree just if \(\APDA\) does \emph{not} accept \(w\)).
Reading \(\ACCEPT\) in state \(q_i\) (with $i<|w|$) on the other hand means that \(\APDA\) does not accept \(w\),
so that the run of \(\trBUCHI{}{w}\) succeeds.
The remaining transition rules are analogous to those of \(\trBUCHI{}{}\).

By the construction above, {\(w\) is \emph{not} accepted by \(\APDA\) if, and only if, the tree generated by \(\tPDA_{\APDA}'\) is accepted by \(\trBUCHI{}{w}\)}. Since  only \(\trBUCHI{}{w}\) depends on the input word \(w\), we get:
\begin{thm}
\label{th:hardness-in-size-of-trivial-APT}
The trivial APT acceptance problem of trees generated by order-$n$ recursion schemes is $n$-EXPTIME-hard in the size of the APT.
\end{thm}


\noindent
To our knowledge, the lower bound (of the complexity of model-checking recursion schemes)
in terms of the size of APT for the
entire
class of APT {is new}.



%% file: disjunctiveAPT.tex
\section{Disjunctive APT and Complexity of Model Checking}
\label{sec:disjunctiveAPT}


\newcommand\pathlang[1]{\mathcal{W}\,{#1}}
\newcommand\vp{\hbox{\bf VP}}
\newcommand\cgraph[1]{\mathit{Gr}_{#1}}




A \emph{disjunctive APT} is an APT whose transition function
\(\delta\) is disjunctive, i.e.~{\(\delta\) maps each state to} a positive boolean formula
{$\theta$} that contains only disjunctions and no conjunctions, as given by the grammar
\(\theta \; ::= \; \TRUE \; | \; \FALSE \; | \; (i, q) \; | \;  \theta \vee \theta \).
Disjunctive APT can be used to describe path (or linear-time) properties of trees.

First we give a logical characterization of disjunctive APT as follows. Call $\cal D$ the following ``disjunctive fragment'' of the modal mu-calculus:
\[\phi, \psi \; ::= \; 
\TRUE 
\; | \; \FALSE
\; | \;
P_f \wedge \phi
\; | \; Z
\; | \; \phi \vee \psi
\; | \; \diaform{i}{\phi}
\; | \; \nuform{Z}{\phi}
\; | \; \muform{Z}{\phi}
\]
where $f$ ranges over symbols in $\Sigma$, and $i$ over $\set{1, \cdots, m}$ where $m$ is the largest arity of the symbols in $\Sigma$.
A proof {of the following proposition} is given in Appendix~\ref{sec:equi-expressivity}.

\begin{prop}[Equi-Expressivity]
\label{prop:equi-expressivity2}
The logic $\cal D$ and 
disjunctive APT are equivalent for defining possibly-infinite
ranked trees. I.e.~for every closed $\cal D$-formula, there is a disjunctive APT that defines the same tree language, and vice versa.
\end{prop}

\begin{remark}
For defining languages of ranked trees, disjunctive APT %
are a proper subset of the \emph{disjunctive formulas} in the sense of Walukiewicz and Janin  \cite{JW95}. 
    For example, the disjunctive formula $(1 \rightarrow \set{{\sf t}}) \wedge (2 \rightarrow \set{{\sf t}})$ is not equivalent to any disjunctive APT.
\end{remark}


In the rest of the section, we show that the model checking problem for order-\(n\) recursion schemes is \((n-1)\)-EXPTIME complete for disjunctive APT.

\subsection{Upper Bound}


{Since our proof is based on Kobayashi and Ong's type system for recursion schemes~\cite{KO09LICS} and relies heavily on the machinery and techniques developed therein, we shall just sketch a proof here; a detailed proof will be presented in the journal version of \cite{KO09LICS}.} An alternative proof, 
{also sketched} but based on \emph{variable profiles}~\cite{Ong06LICS}, is given in Appendix~\ref{app:alt-proof}.

\begin{thm}\label{lem:pdisj}
 Let $\GRAM$ be an order-$n$ recursion scheme and
 $\mathcal{B}$ a disjunctive APT. It {is decidable} in
 $(n-1)$-EXPTIME whether $\mathcal{B}$ accepts the value tree
 $\sem{\GRAM}$. 
\end{thm}


In a recent {paper}~\cite{KO09LICS},
we constructed
an intersection type system {equivalent to the modal mu-calculus model checking of
recursion schemes, in the sense that {for every APT, there is a type system such that}
the tree generated by a recursion scheme is accepted by the APT if, and only if,
the recursion scheme is typable in the type system}.
The model checking problem is thus reduced to a type checking problem.
The main idea of the type system is to refine the tree type \(\T\)
by the states and priorities of an APT. The type \(q\) describes a tree that is accepted
by the APT with \(q\) as the start state.
The type \((\theta_1,m_1)\land (\theta_2,m_2)\ra q\), which refines the type \(\T\ra \T\),
describes a tree function that takes
an argument which has types \(\theta_1\) and \(\theta_2\), and returns
a tree of type \(q\).

The type checking algorithm presented in 
\cite{KO09LICS} is \(n\)-EXPTIME in the combined size of the order-$n$ recursion scheme
and the APT 
({more precisely,\footnote{According to Schewe's recent result~\cite{Schewe07} on the complexity of parity games,
the part \(r^{1+\lfloor m/2 \rfloor}\) can be further reduced to
roughly \(r^{1+m/3}\).} $O(r^{1 + \lfloor m/2 \rfloor} {\bf exp}_n((a \, |Q| \, m)^{1+\epsilon}))$ for \(n\geq 2\),
where $r$ is the number of rules, $a$ is the largest {arity} of the symbols in the scheme, $m$ is the largest priority, and $|Q|$ is the number of states}).
The bottleneck of the algorithm is the number of (atomic) intersection types,
where the set \(\AtomTy{\kappa}\) of atomic types refining a simple type \(\kappa\)
is inductively defined by:
\iffull
\[
\begin{array}{rll}
\AtomTy{\T} &:=& Q \\
\AtomTy{\kappa_1\ra \kappa_2} &:=& \set{\IT S \ra \EAT \mid \EAT\in \AtomTy{\kappa_2}, S\subseteq \AtomTy{\kappa_1}\times P}
\end{array}
\]
\fi
where \(Q\) and \(P\) are the sets of states and priorities respectively.

According to the syntax of atomic types above,
the number of atomic types refining a simple type of order \(n\)
is \(n\)-exponential in general.
In the case of disjunctive APT, however, for each type
of the form \(\T\ra\cdots \ra \T\ra \T\), we need to consider
only atomic types of the form \(\IT S_1 \ra \cdots \ra \IT S_k \ra q\), where
at most one of the \(S_i\)'s is a singleton set and the other \(S_j\)'s are empty.
Intuitively, this is because a run-tree of a disjunctive APT consists of a single path, so that
the run-tree visits only one of the arguments, at most once.
In fact, we can show that,
 if a recursion scheme is typable in
the type system for a disjunctive APT, the recursion scheme is typable
in a {restricted type system in which order-\(1\) types are constrained as described above}: this follows from the proof of completeness of the type
system~\cite{KO09LICS}, along with the property of the accepting run-tree mentioned
above.
Thus, the number of atomic types is \(k\times |Q|\times {|P|}\times |Q|\)
({whereas it is exponential for an arbitrary APT}). Therefore,
the number of atomic types possibly assigned to a symbol of order \(n\) is
\((n-1)\)-{exponential}.
By running the same type checking algorithm as \emph{ibid.}~(but with {order-\(1\) types constrained} as above),
order-$n$ recursion schemes can be type-checked (i.e.~model-checked) in \((n-1)\)-EXPTIME.

\subsection{Lower Bound}
\label{disjunctiveAPT:lower-bound}

We show the lower bound by a reduction of the emptiness problem of the finite-word
language accepted by an order-$n$ deterministic PDA, which is $(n-1)$-EXPTIME complete \cite{Engelfriet91}.

Let \(\APDA\) be an order-$n$ deterministic PDA, given by $\APDA = \langle P, p_0, \Gamma, {\Sigma},
  \delta, F\rangle$ where \(\delta\) is a partial function from
\(P\times ({\Sigma}\cup\set{\epsilon}) \times \Gamma\) to \(P\times \OP{n}\). We shall construct an order-$n$ tree-generating PDA \(\tPDA_{\APDA}{}\),
which simulates all possible input and \(\epsilon\)-transitions of \(\APDA\), and outputs \(\terminal{e}\)
only when \(\APDA\) reaches a final state. 

The order-$n$ PDA \(\tPDA_{\APDA}{}\) is given by:
\[\tPDA_{\APDA}{} =
   \langle {\set{\terminal{e} \mapsto 0} \cup\set{\terminal{br}_{m} \mapsto m \mid 0\leq m\leq N}}, \Gamma,
   {P \cup (P \times \OP{n})}, \delta', p_0\rangle\]
\[
\begin{array}{l}
N = \MAX_{p\in P, \gamma\in\Gamma} |\set{(p',\theta')\mid \exists \alpha\in\Sigma\cup\set{\epsilon}.
                     \delta(p,\alpha,\gamma) = (p',\theta')}|\\
{\delta'}(p,\gamma) =  (\terminal{e}; {\epsilon}) \mbox{ if $p\in F$}\\
{\delta'}(p, \gamma) = (\terminal{br}_m; (p_1,\theta_1),\ldots,(p_m,\theta_m))\\
\qquad   \mbox{ if $p\not\in F$ and
              $\set{(p_1,\theta_1),\ldots,(p_{{m}},\theta_{{m}})} = \set{(p',\theta')\mid \exists \alpha\in\Sigma\cup\set{\epsilon}.
                     \delta(p,\alpha,\gamma) = (p',\theta')}$}\\
{\delta'((p, \theta), \gamma) = (p, \theta)}\\
\end{array}
\]
A state of \(\tPDA_{\APDA}\) is either a state of \(\APDA\) (i.e.~an element of \(P\)), or
a pair \((p, \theta)\). In state \(p\in P\), \(\tPDA_{\APDA}\) constructs a node labeled by {\(\terminal{br}_m\)},
and spawns subtrees for simulating possible input or \(\epsilon\)-transitions of \(\APDA\) from state \(p\).

By a result of Knapik et al.~\cite{Knapik02FOSSACS}, we can construct an
equi-expressive order-$n$ safe recursion scheme \(\GRAM\).
Let \(\GRAM'\) be the recursion scheme obtained from \(\GRAM\) by
(i) replacing each terminal symbol \(\terminal{br}_m \; (m>2)\) with a non-terminal
\(\textit{Br}_m\) of the same arity, and (ii) adding the rule:
\[\textit{Br}_m\,x_1\,\cdots\,x_m \Hra \terminal{br}_2\, x_1\, (\terminal{br}_2\, x_2 (\cdots (\terminal{br}_2\, x_{m-1}\,x_m))).\]
By the construction, the finite word-language accepted by \(\APDA\) is non-empty if, and only if,
the value tree of \(\GRAM'\) has a node labelled \(\terminal{e}\).
\changed{The latter property can be expressed by the following disjunctive APT
\(\trBUCHI{}{}\).} (The purpose of transforming \(\GRAM\) into \(\GRAM'\) was
to make the disjunctive APT independent of \(\APDA\).)
\changed{\[
\begin{array}{l}
\trBUCHI{}{} := \anglebra{ \set{q_0}, \set{\terminal{e}, \terminal{br}_2}, \delta, q_0, \set{q_0\mapsto 1}}\\
\mbox{where }\delta(q_0, \terminal{br}_2) = (1,q_0)\lor (2,q_0)\mbox{ and }
\delta(q_0, \terminal{e}) = \TRUE
\end{array}
\]}
Thus, we have:
\begin{thm}
The disjunctive APT acceptance problem for the tree generated by an 
order-$n$ recursion scheme is $(n-1)$-EXPTIME-hard in the size of the recursion scheme.
\end{thm}


The problem is
\((n-1)\)-EXPTIME hard also in the size of the disjunctive APT.

As above, let \(\APDA = \langle P, p_0, \Gamma, {\Sigma},
  \delta, F\rangle\) be an order-$n$ deterministic PDA for words.
We may assume that the stack alphabet is \(\set{\gamma_0,\gamma_1}\) (as we can encode an arbitrary stack
symbol as a sequence of \(\gamma_0\) and \(\gamma_1\)).

We first define an order-$n$ tree-generating PDA \(\tPDA_{}{}\) by:
\[
\begin{array}{l}
\tPDA_{}{} = \anglebra{\set{\gamma_0,\gamma_1}, \set{\gamma_0,\gamma_1}, \set{q_0,\theta_1,\ldots,\theta_k}, q_0, \delta_{\changed{\tPDA}}}\\
\delta_{\changed{\tPDA}}(q_0, {\gamma_i}) = (\gamma_i; \theta_1, \ldots, \theta_k) \\
\delta_{\changed{\tPDA}}(\theta_i, {\gamma_j}) = (q_0, \theta_i)
\end{array}
\]
where \(\set{\theta_1,\ldots,\theta_k}\) is the set of order-$n$ stack operations.
The role of \(\tPDA_{}{}\) is to generate a tree simulating all the possible changes of the stack top.
Note that \(\tPDA_{}{}\) is independent of \(\APDA\).

Now let us define a disjunctive APT $\dAPT_{\APDA} = \anglebra{{P}, \set{\gamma_0,\gamma_1}, \delta', p_0, \Pfun}$
as follows.
\[
\begin{array}{l}
\delta'(p, \gamma_i) \; =
\; \left\{  
\begin{array}{ll}
\bigvee \set{(j, p') | \exists \alpha.\delta(p, \gamma_i, \alpha)=(p', \theta_j)} \quad &  \mbox{if  $p\not\in F$}\\{\TRUE}  & \mbox{if $p\in F$}\\
\end{array}\right.\\
\Pfun(p) = \changed{1}
\end{array}
\]

The idea of the above encoding is to let \(\dAPT_{\APDA}\) simulate transitions of
\(\APDA\), while extracting information about the stack top
from the tree generated by \(\tPDA_{}{}\).
Let \(\GRAM\) be an order-\(n\) recursion scheme that generates the same tree as \(\tPDA_{}{}\).
By the above construction, the language of \(\APDA\) is non-empty if, and only if,
\(\dAPT_{\APDA}\) accepts the tree generated by \(\GRAM\).
Since the size of \(\GRAM\) does not depend on \(\APDA\), and the size of \(\dAPT_{\APDA}\) is polynomial in
the size of \(\APDA\), we have:
\begin{thm}
\label{th:dAPT-lb-APT}
The disjunctive APT acceptance problem for trees generated by order-$n$ recursion schemes is $(n-1)$-EXPTIME hard in the size of the APT.
\end{thm}



\subsection{Path Properties}
\label{sec:path-properties}

Path properties of $\Sigma$-labelled trees are relevant to program verification, as demonstrated in the application to resource usage analysis in Section~\ref{sec:app}. The \emph{path language} of a $\Sigma$-labelled tree $t$ is
the image of the map $F$, which acts on the elements of the branch
language of $t$ by ``forgetting the argument positions'' i.e.~
\[F \; : \;
\left\{
\begin{array}{lll}
(f_1, d_1) \, (f_2, d_2) \cdots  &\mapsto &f_1 \, f_2 \cdots\\
(f_1, d_1) \cdots (f_n, d_n) \, {f_{n+1}} &\mapsto &f_1 \cdots
f_n \, {f_{n+1}^\omega}.
\end{array}\right.
\]
For example $\set{f\,a^\omega, f\,f\,a^\omega, f\,f\,b^\omega}$ is the path language of the term-tree $f \, a \, (f \, a \, b)$.
Let $\GRAM$ be a recursion scheme. We write \(\pathlang(\GRAM)\) for the \emph{path language} of \(\sem{\GRAM}\). Thus elements of $\pathlang(\GRAM)$ are infinite words over the alphabet $\Sigma$ which is now considered unranked (i.e.~arities of the symbols are forgotten).


\begin{thm}\label{lem:path}
Let $\GRAM$ be an order-$n$ recursion scheme. The following problems are $(n-1)$-EXPTIME complete.

\begin{enumerate}[\em(i)]
\item \(\pathlang(\GRAM)\cap \lang(\WA)\; \stackrel{?}{=}\;\emptyset\), where \(\WA\) is a \emph{non-deterministic} parity word automaton.

\item $\pathlang(\GRAM) \; \stackrel{?}{\subseteq} \; \mathcal{L}(\WA)$, where \(\WA\) is a \emph{deterministic} parity word automaton.
\end{enumerate}\smallskip

\noindent Furthermore, the problem (i) is (\(n-1\))-EXPTIME hard not only in the size of \(\GRAM\) but also in the size of \(\WA\).
\end{thm}


\begin{proof}
\begin{asparaenum}[(i)]
\item Let $\WA = \anglebra{Q, \, \Sigma, \, \Delta, \, \changed{q_I,}\,\Omega}$ be
  a \emph{non-deterministic} parity word automaton, where $\Delta
  \subseteq Q \times \Sigma \times Q$ and $\Omega : Q \longrightarrow
  \set{0, \cdots, p}$. {Let $m$ be the largest arity of the
    symbols in $\Sigma$.} 
(B\"uchi automata are equivalent to parity automata with two priorities.) We have $\pathlang(\GRAM) \cap \mathcal{L}(\WA) \, \neq \, \emptyset$ if, and only if, $\sem{\GRAM}$ is accepted by the APT
$\mathcal{B} =  \anglebra{Q, \, \Sigma, \, \delta, \, \changed{q_I,}\,\Omega}$
where $\delta : Q \times \Sigma
\longrightarrow {\mathsf B}^+({\set{1, \cdots, m}}  \times Q)$ is a \emph{disjunctive} transition function
\[\delta \; : \; (q, f) \; \mapsto \; 
\bigvee \set{(i, p) : 1 \leq i \leq {\Sigma(f)}, (q, f, p) \in \Delta}.\]
It follows from Theorem~\ref{lem:pdisj} that the problem $\pathlang(\GRAM) \cap \mathcal{L}(\WA) \, \stackrel{?}{=} \, \emptyset$ can be decided in $(n-1)$-EXPTIME.

Let \(\WA\) be a parity word automaton that accepts {\(\Sigma^* \, {\terminal{e}}^\omega\)}, and \(\GRAM'\) be the recursion scheme
in Section~\ref{disjunctiveAPT:lower-bound}.
Then, \(\pathlang(\GRAM') \cap \mathcal{L}(\WA) \; \neq \; \emptyset\) if, and only if,
\(\GRAM'\) has a {node labelled \({\terminal{e}}\)}. Thus, the problem
$\pathlang(\GRAM) \cap \mathcal{L}(\WA) \; \stackrel{?}{=} \; \emptyset$
is $(n-1)$-EXPTIME-hard in the size of \(\GRAM\).

\medskip

To show the lower bound in the size of \(\WA\), we modify the construction of \(\tPDA\) and \(\dAPT_{\APDA}\)
as follows.
Let \(\tPDA'\) be the order-\(n\) tree-generating PDA given by:
\[
\begin{array}{l}
\tPDA_{}{} := \anglebra{\set{\gamma_0,\gamma_1,\theta_1,\ldots,\theta_k}, \set{\gamma_0,\gamma_1},
\set{q_0,q_1,\ldots,q_k,\theta_1,\ldots,\theta_k,}, q_0, \delta}\\
\delta(q_0, {\gamma_i}) = (\gamma_i; q_1, \ldots, q_k) \mbox{ for \(0\leq i\leq 1\)}\\
\delta(q_j, \gamma_i) = (\theta_i; \theta_i) \mbox{ for \(0\leq i\leq 1, 1\leq j\leq k\)}\\
\delta(\theta_j, {\gamma_i}) = (q_0, \theta_i)\mbox{ for \(0\leq i\leq 1, 1\leq j\leq k\)}
\end{array}
\]
The difference from \(\tPDA\) is that \(\tPDA'\) outputs not only stack top symbols but also stack operations (which were
coded as branch information in the case of \(\tPDA\)).
Let \(\WA_{\APDA}\) be the non-deterministic parity word automaton given by:
\[
\begin{array}{l}
\WA_{\APDA} := \anglebra{P\cup (P\times \set{0,1}), \set{\gamma_0,\gamma_1}, \delta', p_0, \Pfun}\\
\delta'(p, \gamma_i) = \set{(p, i)}
  \mbox{  if  $p\not\in F$}\\
\delta'((p,i), \theta_j) =
  \set{p' \mid \exists \alpha.\delta(p, \gamma_i, \alpha)=(p', \theta_j)}\\
\delta'(p, \gamma_i) = \set{p}  \mbox{  if  $p\in F$}\\
\delta'(p, \theta_j) = \set{p}\\
\Pfun(p) = \left\{
  \begin{array}{ll}
     2 & \mbox{if $p\in F_{\APDA}$}\\
     1 & \mbox{otherwise}
  \end{array}\right.
\end{array}
\]
Let \(\GRAM\) be a recursion scheme that generates the same tree as \(\tPDA'\).
Then, the language of \(\APDA\) is empty if, and only if, \(\pathlang(\GRAM)\cap \lang(\WA_{\APDA})=\emptyset\).
Since \(\GRAM\) does not depend on \(\APDA\),
$\pathlang(\GRAM) \cap \mathcal{L}(\WA) \; \stackrel{?}{=} \; \emptyset$ is (\(n-1\))-EXPTIME hard also in the size
of \(\WA\).

\medskip

\item Let $\WA$ be a \emph{deterministic} parity word automaton
$\WA =  \anglebra{Q, \Sigma, \delta_{\WA}, q_0, \Omega}$, where $\delta_\WA  :
Q \times \Sigma \longrightarrow Q$ and $\Omega : Q \longrightarrow \set{0, \cdots, p}$.
Define $\overline A = \anglebra{Q, \Sigma,
\delta_{\WA}, q_0, \overline \Omega}$ where $\overline \Omega : q \mapsto
(\Omega(q) + 1)$. Note that because of determinacy,
$\mathcal{L}(\overline {\WA}) = \Sigma^\omega \setminus \mathcal{L}(\WA)$. Now we have
$\pathlang(\GRAM) \subseteq \lang(\WA)$ if, and only if,
$\pathlang(\GRAM) \cap \lang(\overline{\WA}) =\emptyset$.
Thus, the problem  $\pathlang(\GRAM) \; \stackrel{?}{\subseteq} \; \mathcal{L}(\WA)$
is \((n-1)\)-EXPTIME. Moreover, since the language \(\Sigma^* \, {\terminal{e}}^\omega\) is accepted by
a deterministic parity {word} automaton, the problem is also \((n-1)\)-EXPTIME hard (in the size of \(\GRAM\)).
\end{asparaenum}
\end{proof}

\medskip

The decision problems {\textsc{Reachability}} (i.e.~whether \(\sem{\GRAM}\) has a node labelled by a given symbol \(\terminal{e}\))
and {\textsc{Finiteness}} (i.e.~whether \(\sem{\GRAM}\) is finite) are instances of Problem (i) of Theorem~\ref{lem:path}; hence they
are in \((n-1)\)-EXPTIME (the former is \((n-1)\)-EXPTIME complete, by the proof of Section~\ref{disjunctiveAPT:lower-bound}).

Consider the problem \textsc{LTL Model-Checking}: 
\begin{quote}
``Given an LTL-formula
$\phi$ (generated from atomic propositions of the form $P_f$ with $f
\in \Sigma$) and an order-$n$ recursion scheme $\GRAM$, does every
path in $\sem{\GRAM}$ satisfy $\phi$? (Precisely, is $\pathlang(\GRAM)
\subseteq \sem{\phi}$?)''
\end{quote}
As a corollary of Theorem~\ref{lem:path}, we have:
\begin{cor}
\textsc{LTL Model-Checking}
(i.e.~given order-$n$ recursion scheme $\GRAM$ and LTL-formula $\phi$, is $\pathlang(\GRAM) \subseteq \sem{\phi}$?)
is $(n-1)$-EXPTIME complete in the size of $\GRAM$.
\end{cor}
\begin{proof}
The upper bound 
 follows from Theorem~\ref{lem:path}(i):
note that $\pathlang(\GRAM) \subseteq \sem{\phi}$ is equivalent to \(\pathlang(\GRAM)\cap \sem{\neg\phi}=\emptyset\),
and because $\sem{\neg\phi}$ is
$\omega$-regular, it is recognizable \cite{Tho97} by a parity automaton.

The lower bound follows from the $(n-1)$-EXPTIME hardness of
\textsc{Reachability}: checking whether a recursion scheme satisfies
the formula $G (\neg \terminal{e})$
is $(n-1)$-EXPTIME hard in the size of the recursion scheme.
\end{proof}
\noindent
Note however that \textsc{LTL Model-Checking} is \(n\)-EXPTIME
in the size of the LTL-formula \(\phi\),
as the size of the corresponding parity word automaton is exponential in \(\phi\) in general~\cite{VW94}. 

%% file: app.tex
\section{Application to Resource Usage Verification}
\label{sec:app}

Now we apply the result of the previous section to
show that the resource usage verification {problem}~\cite{IK05TOPLAS} is \((n-1)\)-EXPTIME complete.
The aim of resource usage verification is to check whether
a program accesses each resource according to {a given} resource specification.
For example, consider the following program.
\begin{quote}
\begin{verbatim}
let rec g x = if rand() then close(x) else (read(x); g(x)) in
let r = open_in "foo" in g(r)
\end{verbatim}
\end{quote}
Here, \texttt{rand()} returns a non-deterministic boolean.
The program first defines a recursive function \texttt{g} that takes a file pointer \texttt{x} as an argument parameter,
closes it after some read operations. The program then opens a read-only file
{``\texttt{foo}'',} and passes it to \texttt{g}.
For this program, the goal of the verification is to statically check that the file is
eventually closed before the program terminates, and after it is closed, it is never
read from or written to.

Kobayashi~\cite{Kobayashi09POPL}
recently showed that the resource usage verification problem is decidable for the simply-typed \(\lambda\)-calculus
with recursion, generated from a base type of booleans, and augmented by resource creation/access primitives,
by reduction to the model checking problem for recursion schemes. %
Prior to Kobayashi's work~\cite{Kobayashi09POPL}, only sound but incomplete
verification methods have been proposed.

Following \cite{Kobayashi09POPL}, we consider below a simply-typed, call-by-name
 functional language
with only top-level function definitions and resource usage primitives.\footnote{Note that
programs in call-by-value languages can be transformed into
this language by using the standard CPS transformation and \(\lambda\)-lifting.}
A \emph{program} is a triple
 \((D, S, \WA)\) where \(D\) is a set of function definitions,
\(S\) is a function name (representing the main function), and
\(\WA = (Q_\WA,\Sigma_\WA,\delta_\WA,q_{0,\WA}, F_\WA)\) is a deterministic word automaton,
which describes how the state of a resource is changed by each access primitive.
A function definition is of the form \(F\ \seq{x} = e\), where \(e\) is given by:%
\[
\begin{array}{l}
e ::= \unit \mid x \mid F \mid e_1 e_2 \mid
      \ifexp{e_1}{e_2}\mid \nuexp{q}{e} \mid \acc{\f}{e_1}{e_2}
\end{array}
\]
The term \(\unit\) is the unit value.
The term \(\ifexp{e_1}{e_2}\) is a non-deterministic branch between
\(e_1\) and \(e_2\). The term \(\nuexp{q}{e}\) creates
a fresh resource, 
and passes it to \(e\) (which is a function that takes a resource as
an argument). Here, \(q\) represents the initial state of a resource; the automaton \(\WA\) specifies
how the resource should be accessed afterwards: see the operational semantics given later.
The term \(\acc{\f}{e_1}{e_2}\) accesses the resource \(e_1\) with the primitive of name \(\f (\in \Sigma_\WA)\)
and then executes \(e_2\).

Programs must be simply typed; the two base types are \(\Tvoid\) for unit values and \(\Tres\) for resources.
The body of each definition must have type
\(\Tvoid\) (in other words, resources cannot be used as return values; {this} requirement can be
enforced by the CPS transformation \cite{Plotkin75,DanvyF92}). The constants \(\IFNONDET\), \(\NEW{\spec}\),
and {\(\ACC{\f}\)} are given the following types.
\[
\IFNONDET\COL \Tvoid\ra\Tvoid\ra\Tvoid,
\NEW{\spec}\COL (\Tres\ra \Tvoid)\ra \Tvoid,
\ACC{\f}\COL \Tres \ra \Tvoid \ra \Tvoid
\]

\begin{exa}
\label{ex:file}
The program given at the beginning of this section can be expressed as
\((D, S, \WA)\) where
\[
\begin{array}{l}
D = \set{\textit{S} = \nuexp{q_1}(G\ \unit), G\ k\ x = \ifexp{(\acc{c}{x}{k})}{(\acc{r}{x}{(G\ k\ x)})}}\\
\WA = (\set{q_1,q_2}, \set{\terminal{r},\terminal{c}}, \delta,q_1, \set{q_2})\\
\delta(q_1,\terminal{r})=q_1\qquad \delta(q_1,\terminal{c})=q_2
\end{array}
\]
Here, \(G\) corresponds to the function \(g\) in the original program, and the additional parameter
\(k\) represents a continuation.
The automaton \(\WA\) specifies that the resource should be accessed according to \(\terminal{r}^*\terminal{c}\).
\end{exa}

We introduce the operational semantics to formally define the resource usage verification problem.
A run-time state is either an error state \(\ERROR\) or
a pair \((\rst, e)\) where \(\rst\) is a finite map from variables to \(Q_\WA\), which represents the state of each resource.
The reduction relation \(\red{D,\WA}\) on run-time states is defined by:

\infrule{F\ \seq{x} = e' \in D}{(\rst, F\ \seq{e}) \red{D, \WA} (\rst, \subst{\seq x}{\seq e}e')}
\infrule{}{(\rst,\ifexp{e_1}{e_2})\red{D, \WA} (\rst,e_1)}
\infrule{}{(\rst,\ifexp{e_1}{e_2})\red{D, \WA} (\rst,e_2)}
\infrule{x\not\in \dom(\rst)}{(\rst, \nuexp{q}e) \red{D, \WA} (\rst\set{x\mapsto q}, e\, x)}
\infrule{\delta_\WA(q, a)=q'}{(\rst\set{x\mapsto q}, \acc{a}{x}{e}) \red{D, \WA} (\rst\set{x\mapsto q'}, e)}
\infrule{\delta_\WA(q, a)\mbox{ is undefined}}
   {(\rst\set{x\mapsto q}, \acc{a}{x}{e}) \red{D, \WA} \ERROR}

\begin{exa}
Recall the program in Example~\ref{ex:file}.
It can be reduced as follows.
\[
\begin{array}{rcl}
(\emptyset, S)
 &\red{D,\WA}& (\emptyset, \nuexp{q_1}(G\, \unit))\\
 &\red{D,\WA}& (\set{y\mapsto q_1}, G\, \unit\,y)\\
 &\red{D,\WA}& (\set{y\mapsto q_1}, \ifexp{(\acc{c}{y}{\unit})}{(\acc{r}{y}{(G\,\unit\,y)})})\\
 &\red{D,\WA}& (\set{y\mapsto q_1}, \acc{r}{y}{(G\,\unit\,y)})\\
 &\red{D,\WA}& (\set{y\mapsto q_1}, G\,\unit\,y)\\
 &\red{D,\WA}& (\set{y\mapsto q_1}, \ifexp{(\acc{c}{y}{\unit})}{(\acc{r}{y}{(G\,\unit\,y)})})\\
 &\red{D,\WA}& (\set{y\mapsto q_1}, \acc{c}{y}{\unit})\\
 &\red{D,\WA}& (\set{y\mapsto q_2}, {\unit})
\end{array}
\]
\end{exa}

We can now formally define the resource usage verification problem.
\begin{defi}[resource usage verification problem]
\label{df:resource-usage}
A program \((D,S,\WA)\) is \emph{resource-safe} if
(i) \((\emptyset, S)\not\reds{D,\WA} \ERROR\), and (ii) if \((\emptyset,S)\reds{D,\WA} (\rst,\unit)\)
then \(\rst(x)\in F_\WA\) for every \(x\in\dom(\rst)\).
The \emph{resource usage verification} is the problem of checking whether a program is resource-safe.
\end{defi}

\begin{exa}
The program given in Example~\ref{ex:file} is resource-safe.
The program obtained by replacing the body of \(G\)
(i.e. \(\ifexp{(\acc{c}{x}{k})}{(\acc{r}{x}{(G\ k\ x)})}\)) with \(\acc{r}{x}{(G\,k\,x)}\) is also resource-safe;
it does not terminate, so that it satisfies condition (ii) of Definition~\ref{df:resource-usage} vacuously.
The program \(D'\) obtained by replacing the definition of \(G\) with:
\[ G\ k\ x = \ifexp{k}{(\acc{r}{x}{(G\ k\ x)})}\]
is not resource-safe, as \((\emptyset, S) \reds{D',\,\WA} (\set{y\mapsto q_1},\unit)\) and \(q_1\not\in F_\WA\).
\end{exa}

We show below that the resource usage verification is \((n-1)\)-EXPTIME complete for \(n\geq 3\),
where \(n\) is the largest order of types of terms in the source program. Here, the order of a type is defined by:
\[
\order(\Tvoid)=0 \qquad \order(\Tres)=1 \qquad \order(\kappa_1\ra \kappa_2)=\MAX(\order(\kappa_1)+1, \order(\kappa_2))
\]
Note that \(3\) is the
lowest order of a closed program that creates a resource, since \(\NEW{\spec}\) has order \(3\).

The lower-bound can be shown by reduction of the reachability problem for a recursion scheme to
the resource usage verification problem: Given a recursion scheme
\(\GRAM = (\TERMS, \NONTERMS,\RULES,S)\),
let \((D,S,\WA)\) be the program given by:
\[
\begin{array}{l}
D = \set{F\,\seq{x}=\gtop{t} \mid F\,\seq{x}\Hra t\in \RULES}
   \cup \set{\textit{Fail}\ x = \acc{\terminal{fail}}{x}{\unit}}\\
\gtop{F\,t_1\,\cdots\,t_m} = F\,\gtop{t_1}\,\cdots\,\gtop{t_m}\\
\gtop{\terminal{e}} = \changed{\NEW{q}}\, \textit{Fail}\\
\gtop{a\,t_1\,\cdots\,t_m} = \ifexp{\gtop{t_1}}{(\cdots (\ifexp{\gtop{t_{m-1}}}{\gtop{t_m}}))} \mbox{ ($a\neq \terminal{e}$)}\\
\WA = (\set{q}, \set{\terminal{fail}}, \emptyset, q, \set{q})
\end{array}
\]
Then, the value tree of \(\GRAM\) contains \(\terminal{e}\) if and only if
the program \((D,S,\WA)\) is \changed{not} resource-safe.
Since resource primitives occur only in the encoding of \(\terminal{e}\),
the order of the program is
the maximum of \(3\) and the order of the recursion scheme.

\newcommand\qerror{q_{\textbf{error}}}
To show the upper-bound, we transform a program \((D, S,\WA)\) into
a recursion scheme \(\GRAM_{(D,S,\WA)}\), which generates a tree representing all possible
(resource-wise) access sequences of the program~\cite{Kobayashi09POPL}, and
a disjunctive APT \(\dAPT_{(D,S,\WA)}\), which accepts trees containing an invalid resource access sequence,
so that \((D,S,\WA)\) is resource-safe if, and only if, \(\dAPT_{(D,S,\WA)}\) rejects the value tree
of \(\GRAM_{(D,S,\WA)}\).

The recursion scheme \(\GRAM_{(D,S,\WA)}=(\TERMS, \NONTERMS,\RULES,S)\) is given by:
\[
\begin{array}{l}
\TERMS = \set{a\mapsto 1\mid a\in A} \cup \set{\NU{q}\mapsto 2 \mid q\in Q_\WA}
    \cup \set{\unit\mapsto 0, \terminal{i}\mapsto 1, \terminal{k}\mapsto 1, \BR\mapsto 2}\\
\NONTERMS = (\mbox{the set of function symbols in \(D\)})\\
\qquad \cup\set{\IFNONDET\mapsto \T\ra\T\ra\T}
\cup\set{\ACC{a}\mapsto (\T\ra\T)\ra\T\ra\T\mid a\in A}\\
\qquad \cup \set{\nuexp{q}\mapsto ((\T\ra \T)\ra \T)\ra \T \mid q\in Q_\WA}\\
\RULES = \set{F\,\seq{x}\Hra e \mid F\,\seq{x}= e\in D} \\
\qquad  \cup
\set{\ifexp{x}{y} \Hra \terminal{br}\,x\,y, \quad
\acc{\f}{x}{k} \Hra x\,(\f\,k), \quad
\nuexp{q} k \Hra \NU{q} (k\, \terminal{i})\, (k\, \terminal{k})}\\
\end{array}
\]
Here, \(A\) is the set of the names of access primitives that occur in \(D\).

The preceding encoding is slightly different from the one presented in \cite{Kobayashi09POPL}.
The terminal symbol \(\terminal{br}\) represents a non-deterministic choice.
In the rule for \(\NEW{\changed{q}}\), a fresh resource is instantiated to either \(\terminal{i}\) or \(\terminal{k}\)
of arity \(1\). This is a trick used to extract resource-wise
access sequences, by tracking or ignoring the new resource in a non-deterministic manner.
In the first-branch, the resource is instantiated to \(\terminal{i}\), so that all the accesses to the resource
are kept track of. In the second branch, the resource is instantiated to \(\terminal{k}\), so that all the accesses to the resource should
be ignored.
The above transformation preserves types, except that \(\Tvoid\) and
\(\Tres\) are replaced by \(\T\) and \(\T\ra\T\) respectively.

\begin{exa}
\label{ex:acc-tree}
The program in Example~\ref{ex:file} is transformed into the recursion scheme consisting of the following rules:
\[
\begin{array}{rcl}
\textit{S} &\Hra& \nuexp{q_1}(G\ \unit)\\
G\ k\ x &\Hra& \ifexp{(\acc{c}{x}{k})}{(\acc{r}{x}{(G\ k\ x)})}\\
\ifexp{x}{y} &\Hra& \terminal{br}\,x\,y\\
\acc{\f}{x}{k} &\Hra& x\,(\f\,k)\\
\nuexp{\changed{q_1}} k &\Hra& \NU{\changed{q_1}} (k\, \terminal{i})\, (k\, \terminal{k})
\end{array}
\]
Figure~\ref{fig:acc-tree} shows the value tree of the recursion scheme.
The root node represents creation of a new resource (whose initial state is \(q_1\)). The nodes labeled by \(\Tc\) or \(\Tr\)
express resource accesses.
The left and right children are the same, except that each resource access is prefixed by \(\terminal{i}\) in
the left child, while it is prefixed by \(\terminal{k}\) in the right child.
\end{exa}

\begin{figure}
\[
\Tree[ [ [ [ [].{$\unit$} ].{$\Tc$} ].{$\terminal{i}$} [ [ [ [].{$\cdots$} [].{$\cdots$} ].{$\BR$} ].{$\Tr$} ].{$\terminal{i}$} ].{$\BR$}
       [ [ [ [].{$\unit$} ].{$\Tc$} ].{$\terminal{k}$} [ [ [ [].{$\cdots$} [].{$\cdots$} ].{$\BR$} ].{$\Tr$} ].{$\terminal{k}$} ].{$\BR$} ].{$\NU{q_1}$}
\]
\caption{The tree generated by the recursion scheme of Example~\ref{ex:acc-tree}}
\label{fig:acc-tree}
\end{figure}

\begin{exa}
\label{ex:acc-tree2}
Consider the following program, which creates and accesses two resources:
\[
\begin{array}{l}
\textit{S} = \nuexp{q_1}F\\
F\ x = \nuexp{q_1}(G\ \unit\ x)\\
G\ k\ x\ y = \ifexp{(\acc{c}{x}{(\acc{c}{y}{k})})}{(\acc{r}{x}{(\acc{r}{y}(G\ k\ x\ y))})}
\end{array}
\]
It is transformed into the recursion scheme consisting of the following rules:
\[
\begin{array}{rcl}
\textit{S} &\Hra& \nuexp{q_1}F\\
F\ x &\Hra& \nuexp{q_1}(G\ \unit\ x)\\
G\ k\ x\ y &\Hra& \ifexp{(\acc{c}{x}{(\acc{c}{y}{k})})}{(\acc{r}{x}{(\acc{r}{y}(G\ k\ x\ y))})}\\
\ifexp{x}{y} &\Hra& \terminal{br}\,x\,y\\
\acc{\f}{x}{k} &\Hra& x\,(\f\,k)\\
\nuexp{q_1} k &\Hra& \NU{q_1} (k\, \terminal{i})\, (k\, \terminal{k})
\end{array}
\]
Figure~\ref{fig:acc-tree2} shows the value tree of the recursion scheme.
Of the four subtrees whose roots are labeled by \(\BR\), the leftmost subtree represents accesses to both resources \(x\) and \(y\);
in other words, all the accesses to \(x\) and \(y\) are prefixed by \(\terminal{i}\). In the second subtree, only the accesses to \(x\)
are prefixed by \(\terminal{i}\). In the third subtree, only the accesses to \(y\) are prefixed by \(\terminal{i}\), while
in the rightmost subtree, no accesses are prefixed by \(\terminal{i}\).
\end{exa}

\begin{figure}
\[
\Tree
 [
   [ [ [ [ [ [ [].{$\unit$} ].{$\Tc$} ].{$\terminal{i}$} ].{$\Tc$} ].{$\terminal{i}$}
       [ [ [ [ [ [].{$\cdots$} [].{$\cdots$}  ].{$\BR$} ].{$\Tr$} ].{$\terminal{i}$} ].{$\Tr$} ].{$\terminal{i}$}
     ].{$\BR$}
     [ [ [ [ [ [].{$\unit$} ].{$\Tc$} ].{$\terminal{k}$} ].{$\Tc$} ].{$\terminal{i}$}
       [ [ [ [ [ [].{$\cdots$} [].{$\cdots$}  ].{$\BR$} ].{$\Tr$} ].{$\terminal{k}$} ].{$\Tr$} ].{$\terminal{i}$}
     ].{$\BR$}
   ].{$\NU{q_1}$}
   [ [ [ [ [ [ [].{$\unit$} ].{$\Tc$} ].{$\terminal{i}$} ].{$\Tc$} ].{$\terminal{k}$}
       [ [ [ [ [ [].{$\cdots$} [].{$\cdots$}  ].{$\BR$} ].{$\Tr$} ].{$\terminal{i}$} ].{$\Tr$} ].{$\terminal{k}$}
     ].{$\BR$}
     [ [ [ [ [ [].{$\unit$} ].{$\Tc$} ].{$\terminal{k}$} ].{$\Tc$} ].{$\terminal{k}$}
       [ [ [ [ [ [].{$\cdots$} [].{$\cdots$}  ].{$\BR$} ].{$\Tr$} ].{$\terminal{k}$} ].{$\Tr$} ].{$\terminal{k}$}
     ].{$\BR$}
   ].{$\NU{q_1}$}
 ].{$\NU{q_1}$}
\]
\caption{The tree generated by the recursion scheme of Example~\ref{ex:acc-tree2}}
\label{fig:acc-tree2}
\end{figure}

The disjunctive APT \(\dAPT_{(D,S,\WA)} = (\TERMS, Q, \delta, q_I, \Omega)\), which accepts trees
having a path corresponding to an invalid access sequence, is given by:
\newcommand\qsucc{q_{\textbf{succ}}}
\[
\begin{array}{l}
Q = Q_\WA\cup \set{\overline{q} \mid q\in Q_\WA} \cup \set{q_I}\\
\delta(q_I, \f) = \left\{
  \begin{array}{ll}
            (1, q_I)\lor (2, q_I) & \mbox{ if $\f=\BR$}\\
            (1, q)\lor (2, q_I) & \mbox{ if $\f=\NU{q}$}\\
            \FALSE & \mbox{ if $\f=\unit$}\\
            (1, q_I) & \mbox{ otherwise}
  \end{array}\right.\\
\delta(q, \f) \mbox{(where $q\in Q_\WA$)}= \left\{
  \begin{array}{ll}
            (1, q)\lor (2, q) & \mbox{ if $\f=\BR$}\\
            (1, q) & \mbox{ if $\f=\terminal{i}$}\\
            (1, \overline{q}) & \mbox{ if $\f=\terminal{k}$}\\
            (2, q) & \mbox{ if $\f=\NU{q}$}\\
            \FALSE & \mbox{ if $\f=\unit$ and $q\in F_\WA$}\\
            \TRUE & \mbox{ if $\f=\unit$ and $q\not\in F_\WA$}\\
            (1, q') & \mbox{ if $\f\in A$ and $\delta_\WA(q, \f)=q'$}\\
            \TRUE & \mbox{ if $\f\in A$ and $\delta_\WA(q, \f)$ is undefined}\\
  \end{array}\right.\\
\delta(\overline{q}, \f) \mbox{(where $q\in Q_\WA$)} = (1, q)\\
\Omega(q)=1 \mbox{ for every $q\in Q$}
\end{array}
\]
\(\TERMS\) is the same as that of \(\GRAM_{(D,S,\WA)}\).

The APT reads the root of a tree with state \(q_I\), and traverses a tree to find a path corresponding to
an invalid resource access sequence. After reading \(\NU{q}\) in state \(q_I\), the APT either (i) chooses the left branch
and changes its state to \(q\), the initial state of the new resource, tracking accesses to the resource afterwards;
or (ii) chooses the right branch, ignoring accesses to the new resource.
In the mode to track resource accesses (i.e., in state \(q\in Q\)), the APT changes its state according to
resource accesses, except: (i) upon reading \(\terminal{k}\), it skips the next symbol, which represents an access to
a resource not being tracked, (ii) upon reading \(\NU{q}\), it only reads the right branch, ignoring the resource
created by this \(\NU{q}\) (as it is already keeping track of another resource), (iii) upon reading \(a\in A\) such that
\(\WA(q,\f)\) is undefined or reading \(\unit\) when \(q\not\in F_\WA\),
it terminates successfully (as an invalid access sequence has been found), and (iv) upon reading \(\unit\) at state \(q\in F_\WA\),
it aborts (as a path being read was actually a valid access sequence).
The priority function maps every state to \(1\), so that no infinite run (that corresponds to an infinite
execution sequence of the program without any invalid resource access) is considered an accepting run.

From the construction above, we have:
\begin{thm}
\((D,S,\WA)\) is resource-safe if, and only if, the value tree of \(\GRAM_{(D,S,\WA)}\) is not accepted by \(\dAPT_{(D,S,\WA)}\).
\end{thm}
The proof is similar to the corresponding theorem in \cite{Kobayashi09POPL}, hence omitted.\footnote{As mentioned above, the encoding
presented in this article is slightly different from the one in \cite{Kobayashi09POPL}, but the proofs are
{similar: they are tedious} but rather straightforward.}

Note that the order of \(\GRAM_{(D,S,\WA)}\) is the same as that of \(D\).
Thus, as a corollary of the above theorem and Theorem~\ref{lem:pdisj}, we obtain that the resource usage verification is \((n-1)\)-EXPTIME.


%% file: related.tex
\section{Related Work}
\label{sec:related}


\iffull
Our analysis of the lower bound is based on
Engelfriet's earlier work on the complexity of the 
{iterated pushdown automata word acceptance and emptiness problems}, and the results of Knapik et al.~on the relationship between higher-order PDA
and safe recursion schemes.
\fi

The model checking of recursion schemes for the class of trivial APT
has been studied by Aehlig~\cite{Aehlig07} (under the name ``trivial automata'').
He gave a model checking algorithm, but did not discuss its complexity.
For the same class, Kobayashi~\cite{Kobayashi09POPL} showed that
the complexity is linear in the size of recursion schemes, if the {types and automata are fixed}.
For the full modal \(\mu\)-calculus,
Kobayashi and Ong~\cite{KO09LICS} have shown that
the complexity is \(n\)-EXPTIME in the largest arity of symbols in the recursion scheme,
the number of states of the APT, and the largest priory,
but polynomial in the number of the rules of the recursion scheme.

Our encoding of the word acceptance problem of an order-\(n\) alternating PDA into
the model checking problem of an order-\(n\) tree-generating PDA (the construction of
\(\trPDA{\APDA}{w}\) in Section~\ref{sec:ta-hardness})
is similar to Cachat and Walukiewicz's encoding of the word acceptance problem into the reachability game
on a higher-order pushdown system~\cite{Cachat07}. In fact, the tree generated by \(\trPDA{\APDA}{w}\) seems to correspond
to the unravelling of the game graph of the higher-order pushdown system (where the nodes labelled by
\(\EXISTS\) are {Player's positions}, and those labelled by \(\UNIV\) are Opponent's positions).
Thus, \(n\)-EXPTIME-hardness of model checking for trivial APT (in the size of the recursion scheme) would follow also from
\(n\)-EXPTIME hardness of the reachability game on higher-order pushdown systems~\cite{Cachat07,CarayolHMOS08}.


%% file: conc.tex
\section{Conclusion}
\label{sec:conc}
We have considered two subclasses of APT, and shown that the
model checking of an order-$n$ recursion scheme is
\(n\)-EXPTIME complete for trivial APT, and \((n-1)\)-EXPTIME complete
for disjunctive APT, both in the size of the recursion scheme and in the size of the APT.  As an application, we showed that the resource
usage verification problem is \((n-1)\)-EXPTIME complete.  The
lower bound 
for the finiteness problem
(recall Section~\ref{sec:path-properties}) is left as an open problem.


%% file: ack.tex
\paragraph{Acknowledgments}

We would like to thank the {anonymous reviewers} 
for useful comments.  This work was partially supported by Kakenhi 20240001 and EPSRC EP/F036361.  

%% file: Proof-EquiExpressivity.tex
\section{Characterizing trivial APT and disjunctive APT as modal mu-calculus fragments}
\label{apx:APT_mucalculus}
\newcommand\altdep[2]{\mathit{altdep}_{#2}(#1)}

\label{sec:equi-expressivity}

\subsubsection*{From Logic to Automata}

Consider the following set of modal mu-calculus formulas:
\[\phi, \psi \; ::= \; \TRUE
\; | \; \FALSE
\; | \; P_f
\; | \; Z
\; | \; \phi \wedge \psi
\; | \; \phi \vee \psi
\; | \; \diaform{i}{\phi}
\; | \; \nuform{Z}{\phi}
\; | \; \muform{Z}{\phi}
\]
This is a superset of the fragments \(\mathcal{S}\) and \(\mathcal{D}\) introduced in Sections~\ref{sec:ta-hardness}
and \ref{sec:disjunctiveAPT}, respectively.

We can apply the translation of Kupferman et al.~\cite{KVW00} to a modal mu-calculus formula to get
an equivalent alternating parity tree automaton. 
We just need to modify the definition of \(\delta\) (\cite[page 339]{KVW00}) by:
\[\begin{array}{l}
\delta(P_f, g) = \left\{\begin{array}{ll}
    \TRUE & \mbox{ if $g=f$}\\
 \FALSE & \mbox{ if $g\neq f$}\end{array}\right.\\
\delta(\diaform{i}{\phi}, g) = \left\{\begin{array}{ll}
    \mathit{split}(i, \phi) & \mbox{ if $1\leq i\leq \Sigma(g)$}\\
    \FALSE & \mbox{ otherwise}\end{array}\right.
\end{array}
\]
It is easy to see that the translation maps a \(\mathcal{S}\)-formula to a trivial automaton, and
a \(\mathcal{D}\)-formula to a disjunctive automaton.

\subsubsection*{From Automata to Logic}
Our presentation here follows Walukiewicz \cite{Wal01}. 
Fix an APT ${\cal A} = \anglebra{\Sigma, Q, \delta, \lochanged{q_1}, \Omega}$ where $Q = \set{q_1, \cdots, q_n}$. 
Suppose the ordering $q_1, \cdots, q_n$ satisfies \lochanged{$\Omega(q_i) \geq \Omega(q_j)$} for every $i < j$. Consider the following $n$-tuple of modal mu-calculus formulas---call it $\chi_{\cal A}$---simultaneously defined by least and greatest fixpoints:
\newcommand\colvec[2]{\left( \begin{array}{c}#1\\ \vdots\\ #2\\\end{array}\right)}
\[ \sigma_1 \colvec{Z_{11}}{Z_{1n}} \, . \cdots . \, \sigma_n \colvec{Z_{n1}}{Z_{nn}} \, .\, \colvec{\chi_1}{\chi_n} \]
where $\sigma_i := \mu$ if $\Omega(q_i)$ is odd, and $\nu$ otherwise. For each $1 \leq i \leq n$
\[\chi_i \; := \; \bigvee_{f \in \Sigma} (P_f \wedge \encode{\delta(q_i, f)}).\]
We define \(\encode{\delta(q_i, f)}\) by:
\[ 
\begin{array}{l}
\encode{(d, q_i)} \; := \; \diaform{d}{Z_{ii}}\\
\encode{\TRUE} \; := \; \TRUE\\
\encode{\FALSE} \; := \; \FALSE\\
\encode{\varphi_1\land\varphi_2} \; := \; \encode{\varphi_1}\land\encode{\varphi_2} \\
\encode{\varphi_1\lor\varphi_2} \; := \; \encode{\varphi_1}\lor\encode{\varphi_2} 
\end{array}
\]
\nk{Here, the translation for \(\TRUE\) and \(\FALSE\) is a bit awkward as we do not have \(\TRUE\) and \(\FALSE\) in
\(\mathcal{S}\) and \(\mathcal{D}\).}
\lo{This is a matter of taste, but I would (i) when defining the superset of fragments of $\cal S$ and $\cal D$ at the beginning, include $\TRUE$ and $\FALSE$ and mention that they are shorthand for $\nu Z.Z$ and $\mu Z.Z$ (ii) when defining $\encode{\delta(q_i, f)}$ above, just define $\encode{\TRUE} := \TRUE$, etc.}


Write $\pi_i(\chi_{\cal A})$ to be a modal mu-calculus formula (semantically) equivalent to $\chi_{\cal A}$ projected onto the $i$-th component (which is well-defined by an application of the Beki\u{c} Principle).

Let $\cal A$ be an APT and $t$ a $\Sigma$-labelled ranked tree. Walukiewicz \cite{Wal01}~has shown that $t$ is accepted by $\cal A$ if, and only if, it satisfies $\pi_1(\chi_{\cal A})$ at the root.
\nk{Is this really OK? What do you mean by \(\pi_I\)?} \lo{Sorry, it should be $\pi_1$. It was written $\pi_I$ because the initial state of the APT was wrongly written as $I$ (which I have also corrected).} \nk{ (i) Walukiewicz's theorem is for \(I=1\). 
(ii) In his paper, the acceptance condition is that the \emph{least} priority visited infinitely often is even,
while our acceptance condition is about the \emph{largest} one.} \lo{Yes, as per your email, the ordering between states should be reversed.}

\begin{proposition}
\begin{asparaenum}[(i)]
\item If $\cal A$ is a trivial APT, then $\pi_1(\chi_{\cal A})$ is a $\cal S$-formula.
\item If $\cal A$ is a disjunctive APT, then $\pi_1(\chi_{\cal A})$ is a $\cal D$-formula.
\end{asparaenum}
\end{proposition}


\begin{proof}

(i): If $\cal A$ has only one priority 0, then it follows from the definition that $\chi_{\cal A}$ is constructed using only $\nu$-fixpoint operator. (ii) Since $\delta(q_i, f)$ is a disjunctive formula, 
it follows that every conjunction subformula of {$\chi_{\mathcal{A}}$} is of the form $P_f \wedge \phi$.
\end{proof}

%% file: upperbound-disjunctiveAPT.tex
\section{Alternative Proof of $(n-1)$-EXPTIME Upper-Bound for Disjunctive APT}

\label{app:alt-proof}
We {sketch} an alternative proof of Lemma~\ref{lem:pdisj}, using Ong's \emph{variable profiles}~\cite{Ong06LICS}.

 In order to appreciate the proof sketched below, some knowledge of the workings of a traversal simulating APT is required. In particular it is necessary to know about \emph{variable profiles} and how they are employed.

Since $\mathcal{B}$ is disjunctive, it has an accepting run-tree on $\sem{G}$ just in case it has an accepting run-tree that does not branch (i.e. each node of the run-tree has at most one child). It follows that $\mathcal{B}$ has an accepting traversal tree if and only if it has an accepting traversal tree that does not branch.

The key observation is that the traversal-simulating APT $\mathcal{C}$ thus need only `guess' one exit point when it reaches a node labelled by a variable of order one, even if its type has arity greater than one. It follows that we can simplify the definition of variable profiles. A profile of a ground-type variable has the shape $(x, q, m, \emptyset)$ where $q$ is a state and $m$ a colour, which is the same as the general case. However a profile of a variable $\phi$ of a first-order type $\underbrace{o \rightarrow \cdots \rightarrow o}_k \rightarrow o$ now has the shape $(\phi, q, m, c)$ where $c$ is either empty or a \emph{singleton} set consisting of a profile of a ground-type variable, as opposed to a {set} of such profiles.
The profiles of variables of order two or higher are defined as in the general case. Thus the number of variable profiles of a given order (at least one) is reduced by one level of exponentiation compared to the general case. Now viewing $\vp(A)$ as denoting the set of variable profiles of type $A$ (of order at least one) restricted to containing either empty or singleton interfaces:
\[
 \sum_{A \ \mathrm{ order} \ i \ \mathrm{type}} |\vp(A)| \; = \; O(exp_{i-1}(|\cgraph{G}| \times |Q| \times p))
\]
where $Q$ is the state space of $\mathcal{B}$, $p$ is the number of {priorities}, {and $\cgraph{G}$ is the (finite) graph that unravels to the computation tree $\lambda(G)$}. The number of nodes in the parity game induced by the traversal-simulating APT $\mathcal{C}$ and the computation tree $\lambda(G)$ will thus also have bound
$O(exp_{n-1}(|\cgraph{G}| \times |Q| \times p))$ and using
Jurdzi\'nski's algorithm \cite{Jur00} we have it that the acceptance parity game
can be solved in time $ O(exp_{n-1}(|\cgraph{G}| \times |Q| \times
p))$. The problem thus lies in $(n-1)$-EXPTIME.




%% file: Main.bbl
\begin{thebibliography}{10}

\bibitem{Aehlig07}
K.~Aehlig.
\newblock A finite semantics of simply-typed lambda terms for infinite runs of
  automata.
\newblock {\em Logical Methods in Computer Science}, 3(3), 2007.

\bibitem{BO07}
W.~Blum and C.-H.~L. Ong.
\newblock Safe lambda calculus.
\newblock In {\em Proceedings of the 8th International Conference on Typed
  Lambda Calculi and Applications (TLCA07)}, pages 39--53. Springer-Verlag,
  2007.
\newblock LNCS 4583.

\bibitem{Cachat07}
T.~Cachat and I.~Walukiewicz.
\newblock The complexity of games on higher order pushdown automata.
\newblock {\em CoRR}, abs/0705.0262, 2007.

\bibitem{CarayolHMOS08}
A.~Carayol, M.~Hague, A.~Meyer, C.-H.~L. Ong, and O.~Serre.
\newblock Winning regions of higher-order pushdown games.
\newblock In {\em LICS}, pages 193--204, 2008.

\bibitem{DanvyF92}
O.~Danvy and A.~Filinski.
\newblock Representing control: A study of the cps transformation.
\newblock {\em Mathematical Structures in Computer Science}, 2(4):361--391,
  1992.

\bibitem{Engelfriet91}
J.~Engelfriet.
\newblock Iterated stack automata and complexity classes.
\newblock {\em {Information and Computation}}, 95(1):21--75, 1991.

\bibitem{IK05TOPLAS}
A.~Igarashi and N.~Kobayashi.
\newblock Resource usage analysis.
\newblock {\em {ACM Transactions on Programming Languages and Systems}},
  27(2):264--313, 2005.

\bibitem{JW95}
D.~Janin and I.~Walukiewicz.
\newblock Automata for the modal mu-calculus and related results.
\newblock In {\em Proc.~MFCS}, pages 552--562, 1995.

\bibitem{Jur00}
M.~Jurdzi\'nski.
\newblock Small progress measures for solving parity games.
\newblock In {\em Proc.~STACS}, volume 1770 of {\em Lecture Notes in Computer
  Science}, pages 290--301, 2000.

\bibitem{Knapik02FOSSACS}
T.~Knapik, D.~Niwinski, and P.~Urzyczyn.
\newblock Higher-order pushdown trees are easy.
\newblock In {\em FoSSaCS 2002}, volume 2303 of {\em {Lecture Notes in Computer
  Science}}, pages 205--222. Springer-Verlag, 2002.

\bibitem{KNU02}
T.~Knapik, D.~Niwi{\'n}ski, and P.~Urzyczyn.
\newblock Higher-order pushdown trees are easy.
\newblock In {\em FOSSACS'02}, pages 205--222. Springer, 2002.
\newblock LNCS Vol.~2303.

\bibitem{Kobayashi09POPL}
N.~Kobayashi.
\newblock Types and higher-order recursion schemes for verification of
  higher-order programs.
\newblock In {\em {Proceedings of {ACM} SIGPLAN/SIGACT Symposium on Principles
  of Programming Languages}}, pages 416--428, 2009.

\bibitem{KO09LICS}
N.~Kobayashi and C.-H.~L. Ong.
\newblock A type system equivalent to the modal mu-calculus model checking of
  higher-order recursion schemes.
\newblock In {\em Proceedings of LICS 2009}, pages 179--188. IEEE Computer
  Society Press, 2009.

\bibitem{KVW00}
O.~Kupferman, M.~Y. Vardi, and P.~Wolper.
\newblock An automata-theoretic approach to branching-time model checking.
\newblock {\em J. ACM}, 47(2):312--360, 2000.

\bibitem{Ong06LICS}
C.-H.~L. Ong.
\newblock On model-checking trees generated by higher-order recursion schemes.
\newblock In {\em LICS 2006}, pages 81--90. IEEE Computer Society Press, 2006.

\bibitem{Plotkin75}
G.~D. Plotkin.
\newblock Call-by-name, call-by-value and the lambda-calculus.
\newblock {\em Theor. Comput. Sci.}, 1(2):125--159, 1975.

\bibitem{Schewe07}
S.~Schewe.
\newblock Solving parity games in big steps.
\newblock In {\em Proceedings of FSTTCS 2007}, volume 4855 of {\em {Lecture
  Notes in Computer Science}}, pages 449--460. Springer-Verlag, 2007.

\bibitem{Tho97}
W.~Thomas.
\newblock Languages, automata and logic.
\newblock In G.~Rozenberg and A.~Salomaa, editors, {\em Handbook of Formal
  Languages}, volume~3. Springer-Verlag, 1997.

\bibitem{VW94}
M.~Y. Vardi and P.~Wolper.
\newblock Reasoning about infinite computations.
\newblock {\em Information and Computation}, 115:1Ð37, 1994.

\bibitem{Wal01}
I.~Walukiewicz.
\newblock Pushdown processes: games and model-checking.
\newblock {\em Information and Computation}, 157:234--263, 2001.

\end{thebibliography}
